\documentclass[12pt]{article}
\usepackage[margin=2 cm]{geometry}
\usepackage{comment}
\usepackage{multirow}
\usepackage{graphicx}
\usepackage[font={small,it}]{caption}
\usepackage{mwe}
\usepackage[nottoc]{tocbibind}
\usepackage{amsmath,amssymb,extarrows,mathtools,graphicx,subfigure,setspace}
\usepackage{cite}
\usepackage{tikz}\usetikzlibrary{calc}
\usepackage{braket}
\usepackage{epsfig}
\usepackage[section]{placeins}
\usepackage{slashed}
\usepackage{color}
\usepackage{caption}
\usepackage{amsmath}
\usepackage{hyperref}
\makeatother

\newcommand{\be}{\begin{equation}}
\newcommand{\bea}{\begin{eqnarray}}
\newcommand{\eea}{\end{eqnarray}}
\newcommand{\ba}{\begin{array}}
\newcommand{\ea}{\end{array}}
\newcommand{\ee}{\end{equation}}
\newcommand{\bes}{\begin{equation*}}
\newcommand{\beas}{\begin{eqnarray*}}
\newcommand{\eeas}{\end{eqnarray*}}
\newcommand{\bas}{\begin{array*}}
\newcommand{\eas}{\end{array*}}
\newcommand{\ees}{\end{equation*}}

\numberwithin{equation}{section}

\begin{document}

\onehalfspacing
\vfill
\begin{titlepage}
\vspace{10mm}

\begin{center}

\vspace*{10mm}
\vspace*{1mm}
{\Large  \textbf{The spectral gap of the ABJM model: A holographic perspective from uplifted higher-dimensional geometries}} 
 \vspace*{1cm}
 
{$\text{Mahdis Ghodrati}^{a,b}$}

\vspace*{10mm}

{ \textsl{ $^a $ Michigan Center for Theoretical Physics, Randall Laboratory of Physics,
University of Michigan, Ann Arbor, MI 48109-1040, USA}} 
 \vspace*{0.4cm}

{ \textsl{ $^b$ School of Physics, Institute for Research in Fundamental Sciences (IPM)
P.O. Box: 19395-5531, Tehran, Iran}}

 \vspace*{0.7cm}

\textsl{e-mail: {\href{ghodrati@umich.edu}{ghodrati@umich.edu}}}
 \vspace*{2mm}

\vspace*{1.7cm}

\end{center}

\begin{abstract}
We study the $U(1)^4$ charged black brane solution of four-dimensional gauged supergravity and analyze the fermionic response in this geometry, which is holographically dual to $3d$ $\mathcal{N}=2$ SCFT ABJM models. Similar to \cite{DeWolfe:2013fha}, we show that a gap exists in the states of the conformal field theory, which corresponds to the different limiting behaviors of the two unequal chemical potentials of the four-charge geometry. We study the behavior of this gap and also the stability of the near-horizon geometry by changing the parameters of the theory in various orders. We then categorize the $56$ fermion modes of the geometry, find the coefficients of the Dirac equations for each mode, and comment on the behavior of the solution of the Dirac equation in different near-horizon limits. We then uplift the geometry to five-dimensional and then to eleven-dimensional geometries and, similar to \cite{Fareghbal:2008dy}, we show that in both cases, a piece of $\mathrm{BTZ} \times \mathrm{S}^2$ or $\mathrm{AdS}^3 \times \mathbb{R}^2$ emerges, and as a result, a decoupling sector in the field theory exists. We study the singularity of the near-horizon $4d$ geometry and the mentioned gap in these uplifted geometries.
 \end{abstract}

\end{titlepage}

\tableofcontents

\section{ Introduction }
One method to study strongly coupled systems and characterize them as Fermi liquids, marginal Fermi liquids, or non-Fermi liquids (such as high-$\text{T}_c$ superconductors or strange metals \cite{Hartnoll:2008kx,Li:2011aa}) is to use the gauge/gravity correspondence. This relates a strongly coupled field theory in $d$ dimensions to a weakly coupled gravity theory in one dimension higher \cite{Maldacena:1997re,DeWolfe:2012uv,DeWolfe:2013fha,DeWolfe:2011ts,DeWolfe:2014ifa, Weidner:2006rp}.

In one approach, one first uses the symmetries of the desired field theory to construct a Lagrangian in the gravity theory and then solves the equations of motion to find the characteristics of the field theory side. This is the ``bottom-up'' approach, which has been used extensively. However, this approach can yield unphysical or unstable solutions if the parameters of the theory are chosen arbitrarily (see \cite{Ghodrati:2014spa} for an example). One way to resolve this issue is to use the ``top-down'' approach, which makes a direct connection with string theory and supergravity by choosing a supergravity solution for the bulk geometry from the beginning. This has been used in, among others, \cite{DeWolfe:2012uv,DeWolfe:2013fha,DeWolfe:2011ts}.

In this paper, we choose the extremal four-dimensional $\text{U}(1)^4$ charged black brane solution of $\mathcal{N}=2$ gauged supergravity to holographically study a zero-temperature, non-zero density ABJM system of $3d$ $\mathcal{N}=2$ superconformal field theory with four distinct chemical potentials. These black holes were first introduced in \cite{Cvetic:1999xx}. The main characteristic of these black holes/branes is that they have four charges. For the general extremal form of the theory, the near-horizon geometry is non-singular, the entropy is non-zero at zero temperature, and the near-horizon geometry has an $\mathrm{AdS}_2$ sector.

For simplicity, one can imagine that three of these four charges are equal, and therefore call the system a regular extremal ``3+1-charge'' black brane. One can choose one of these charges to be zero and end up with a ``3-charge'' black brane, which has a singular near-horizon and zero entropy at zero temperature---which is physically more desirable. As mentioned in \cite{DeWolfe:2013fha}, this singularity is of a good type, as it will be resolved in the one-dimension-higher uplifted theory. We check that for both cases---$4d$ $\text{U}(1)^4$ gauged supergravity uplifted to $5d$ and $11d$, and $3d$ $\text{U}(1)^3$ gauged supergravity uplifted to $4d$ and $10d$---the uplifted near-horizon geometries have an $\mathrm{AdS}_3$ or BTZ sector and the singularity is resolved in the uplifted backgrounds.

In \cite{Fareghbal:2008dt} and \cite{Fareghbal:2008dy}, for the $5d$ $\text{U}(1)^3$ and $4d$ $\text{U}(1)^4$ cases respectively, the authors showed that the near-horizon geometry is a decoupled sector, and therefore the dual field theory correspondingly has a decoupled sector where the dynamics there is separated from the rest of the states of the theory. We comment on the relationship between this sector and the gap seen in \cite{DeWolfe:2013fha}, which is related to the different behaviors of the chemical potentials when taking the limits in two different orders: first the near-horizon limit and then the 3+1-charge to 3-charge limit ($q^\prime \to 0$), or the reverse order. In fact, this gap in the CFT is the dual of the decoupled near-horizon $\mathrm{AdS}_3$ or BTZ geometry of the $\text{U}(1)^4$ supergravity black hole in the bulk, first found in \cite{Fareghbal:2008dt}.

We also find that there are differences in the physical nature of the three equal charges and the other unequal charge in the 3+1-charge black brane system. For instance, we show that $k_{\text{osc}}$ can exist in the near-horizon of extremal 3-charge black branes, so the near-horizon can be unstable due to the effects of these charges. But for the 1-charge black brane, $k_{\text{osc}}$ cannot exist, and even with a large 1-charge, the near-horizon would always be stable. Another difference between these two kinds of charges is that the gap can only exist when the 1-charge $q^\prime$ is turned off. Turning on even a small value of $q^\prime$ makes the gap disappear \cite{DeWolfe:2013fha}, which results from non-uniformity of states and a discontinuity in the chemical potential $\mu_1$. Also, in the uplifted geometry, only $q^\prime$ is associated with the Kaluza--Klein charge and compact momentum. The source of the 1-charge gauge field in one dimension higher is the graviphoton, while the source of the 3-charge gauge field $A_\mu$ can be understood by uplifting to $10d$ or $11d$, where it becomes a two-form.

To further study the theory holographically, one can examine the linear response of probe spin-$1/2$ fermions in the background of the $4d$ supergravity charged black brane by solving the Dirac equation for each mode, as has been done for the $5d$ case in \cite{DeWolfe:2012uv,DeWolfe:2011ts,Ahn:2001by}. The authors mention that the main feature of the fermionic response in this top-down approach is that the mass term couples to the running dilaton and diverges at the singularity ($r \to 0$). This divergence of the fermion mass term at the horizon could be a feature of top-down solutions arising from supergravity, which we aim to study here for the four-dimensional case as well.

We categorize the 56 fermion modes of our theory. Then, by embedding the theory in $4d$ $\mathcal{N}=8$ supergravity \cite{DeWit:1982td}, we find the coefficients of the Dirac equations. We then examine the behavior of the Dirac equation and its solutions in the two limits and, therefore, in the different sectors of the theory. Specifically, finding the Fermi surfaces using the Green's function, and the oscillation modes around them, would be of interest. In \cite{DeWolfe:2012uv} and \cite{DeWolfe:2013fha}, all the Fermi surfaces for the 32 modes out of 64 modes of the $5d$ $U(1)^3$ black brane theory that do not mix with the gravitini were found. A similar method could be used in the $4d$ $U(1)^4$ theory for categorizing the fermion modes, solving the Dirac equations, and finding the Fermi surfaces, which is being presented here. It has been mentioned in \cite{DeWolfe:2013fha} that for the extremal 2-charge case, there exists a gap of the order of the chemical potential where fluctuations of the modes inside it are stable and $\Gamma / \omega_{\ast} \to \text{constant}$, corresponding to non-Fermi liquid behavior. We show that such a gap exists in the four-dimensional case as well and similarly check its characteristics by studying the poles of the retarded Green's functions.

\section{ Charged black brane solution }
The Lagrangian of the $4d$ $U(1)^4$ gauged supergravity solution without axions is \cite{Cvetic:1999xx}
\begin{align} \label{eq:lag}
e^{-1} \mathcal{L}_4
&= R-\frac{1}{2}(\partial \vec{\varphi})^2
+8g^2 \bigl(\cosh\varphi_1+\cosh\varphi_2+\cosh\varphi_3\bigr)
-\frac{1}{4}\sum_{i=1}^4 e^{\vec{a}_i\cdot\vec{\varphi}} \Big(F^i_{(2)}\Big)^2,
\end{align}
where
\begin{align} \label{eq:rel}
\vec{\varphi}&=(\varphi_1,\varphi_2,\varphi_3), \quad
\vec{a}_1=(1,1,1), \quad
\vec{a}_2=(1,-1,-1), \quad
\vec{a}_3=(-1,1,-1), \quad
\vec{a}_4=(-1,-1,1).
\end{align}

The special sector of this theory that we are interested in is the one in which three of the four gauge fields are equal to each other. Checking the equations of motion for consistency of the theory implies that in this case three of the four scalar fields should also be set equal to each other. Using $X_1 X_2 X_3 X_4=1$ and \eqref{eq:rel}, we find that $\varphi_1=\varphi_2=-\varphi_3=\varphi$. Then, by changing the parameter to $\varphi=\frac{\phi}{\sqrt{3}}$, the Lagrangian of the 3+1-charge sector becomes
\begin{equation} \label{eq:lagmain}
\mathcal{L}_{(3+1)}
= R-\frac{1}{2}(\partial \vec{\phi})^2
+24g^2 \cosh\frac{\phi}{\sqrt{3}}
-\frac{3}{4} e^{\frac{\phi}{\sqrt{3}}} F^2
-\frac{1}{4} e^{-\sqrt{3}\phi} f^2.
\end{equation}
By setting $F=F'=0$ and taking a constant dilaton $\phi=0$, corresponding to the $\mathrm{AdS}_4$ theory, we can find the coupling constant of the Lagrangian, i.e., $g=\frac{1}{2L}$. One can also find that $X_1=X_2=X_3=e^{-\frac{\phi}{2\sqrt{3}}}$ and $X_4=e^{\frac{\sqrt{3}\phi}{2}}$. The dual of this theory is a $\mathrm{CFT}_3$ with the symmetry group $\mathrm{SO(8)}$, and in our case, with the subgroup $U(1)^4$.

For comparison, the Lagrangian of the $5d$ supergravity theory is \cite{DeWolfe:2013fha}
\begin{equation}
e^{-1} \mathcal{L}
= R-\frac{1}{2}(\partial\varphi)^2
+\frac{8}{L^2}e^{\frac{\varphi}{\sqrt{6}}}
+\frac{4}{L^2} e^{-\frac{2\varphi}{\sqrt{6}}}
-e^{-\frac{4\varphi}{\sqrt{6}}}f_{\mu\nu}f^{\mu\nu}
-2e^{\frac{2\varphi}{\sqrt{6}}}F_{\mu\nu}F^{\mu\nu}
-2\epsilon^{\mu\nu\rho\sigma\tau}f_{\mu\nu}F_{\rho\sigma}A_{\tau},
\end{equation}
whose dual theory is a $\mathrm{CFT}_4$ with symmetry group $\mathrm{SO(6)}$ and subgroup $U(1)^3$. The Chern--Simons term is absent for the $4d$ case \cite{Aharony:2008ug}.

The general form of the static $U(1)^4$ gauged black hole solution in $d=4$ is
\begin{equation} \label{metric}
ds^2=-H^{-\frac{1}{2}} f(r)\, dt^2
+H^{\frac{1}{2}}\left(\frac{dr^2}{f(r)}+r^2 d\Omega_{2,k}^2\right),
\end{equation}
where
\begin{align}
H&=H_1H_2H_3H_4,\\
H_I&=1+\frac{\mu \sinh^2\beta_I}{k\, r}=1+\frac{q_I}{r},\\
f(r)&=k-\frac{\mu}{r}+\frac{r^2}{L^2}H,
\end{align}
with $q_I=\frac{\mu \sinh^2 \beta_I}{k}$ being a parameter of the solution, and $k$ can be $1$, $0$, or $-1$ for spherical, flat, or hyperbolic $(S^2, T^2, H^2)$ foliating the transverse space \cite{Cvetic:1999xx, Fareghbal:2008dy, Duff:1999gh}. Here we consider the case $k=0$, corresponding to a black brane solution, as we want to study a dual non-compact CFT on flat space at the boundary \cite{Seiberg:1996nz}.

The black brane solutions of both the $4d$ and $5d$ theories belong to the family of extremal vanishing horizon (EVH) black holes. In these black holes, for the extremal case where $T \to 0$, $A_h \to 0$, so $S \to 0$, while the ratio $\frac{A_h}{T}$ remains constant. This is a desirable feature, as the physical origin of non-zero entropy at zero temperature is not well understood. In their near-horizon limit, these black hole solutions contain an $\mathrm{AdS}_3$ throat \cite{Fareghbal:2008dt, SheikhJabbari:2011uu, deBoer:2011mm, deBoer:2012fi}.

The gauge fields and scalar fields of the theory are
\begin{gather}
A_I=\frac{Q_I}{q_I}\left(\frac{1}{H_I(r)}-\frac{1}{H_I(r_H)}\right)dt
=\frac{Q_I}{q_I+r_H} \left(1-\frac{q_I+r_H}{q_I+r}\right)dt, \quad
X_I=\frac{H^\frac{1}{4}}{H_I}.
\end{gather}

The difference between these gauge fields and the one in \cite{Fareghbal:2008dy} is that we chose a gauge such that at $r=r_H$, $A_t=0$, so the chemical potentials become finite at the boundary. Also, the $X_I$'s, which parameterize the three scalars, satisfy the relation $X_1 X_2 X_3 X_4 =1$.

For the case of $k=0$, by rescaling $\sinh^2 \beta_I \to k \sinh^2 \beta_I$ and sending $k$ to zero \cite{Cvetic:1999xx}, one gets
\begin{equation}
A_{(1)}^I=\frac{H_I^{-1}(r)-H_I^{-1}(r_H)}{\sinh\beta_I} dt.
\end{equation}
In the above equation, $\sinh^2 \beta_I = \frac{q_I k}{q_I+\mu}$, where $q_I$ and $\mu$ are the parameters of the solution. These five parameters are related to each other by \cite{Fareghbal:2008dy}
\begin{equation}
Q_I=\sqrt{q_I (q_I+\mu)}.
\end{equation}
The physical observables of these $4d$ black hole solutions are the ADM mass and the electric charges \cite{Fareghbal:2008dy}
\begin{gather}
M=\frac{1}{2G_N^{(4)}}(2\mu+q_1+q_2+q_3+q_4), \quad
J_I=\frac{L}{2G_N^{(4)}} Q_I.
\end{gather}
At the horizon, $f(r_H)=0$, so
\begin{equation}\label{eq:exsol}
-\mu r +\frac{1}{L^2}(r+q_1)(r+q_2)(r+q_3)(r+q_4) \Big|_{r=r_H}=0.
\end{equation}
We want to study the 3+1-charge case where
\begin{equation}
q_1=q_2=q_3=q, \qquad q_4=q^{\prime},
\end{equation}
corresponding to
\begin{equation}
Q_1=Q_2=Q_3=Q, \qquad Q_4=Q^{\prime}.
\end{equation}
The prime is used to define a new parameter here. So from \eqref{eq:exsol},
\begin{equation} \label{eq:mu}
\mu =\frac{ (q+r_H)^3 (q'+r_H)}{L^2 r_H}.
\end{equation}
Then, for the general regular case, using \eqref{eq:mu}, the horizon temperature is
\begin{equation}
T_H=\frac{1}{4\pi} \frac{\partial g_{tt}}{\partial r}\Big|_{r=r_H}
=\frac{| 3 r_H^2+2 q' r_H- q q' |}{4 L^2 \pi r_H}
\sqrt{\frac{q+r_H}{q'+r_H}}.
\end{equation}
For the regular extremal 3+1-charge black hole, where $T_H(r_H)=0$, there are two solutions:
\begin{gather} \label{eq:extg}
\mathrm{I.} \quad r_H=-q=Q,\ \text{with any } q^\prime, \nonumber\\
\mathrm{II.} \quad 3 r_H^2+2 q' r_H-q q'=0.
\end{gather}
For the first solution, $\mu=0$ and the solution is BPS. As these black holes are solutions of $11d$ supergravity, a BPS solution with $n$ non-zero charges $q_I$ preserves $32/2^n$ supersymmetries. For the second solution, $\mu=\frac{(q+r_H)^4}{L^2 (q-2r_H)}$, and for $q>2r_H$, $\mu>0$ and the solution is non-BPS with no supersymmetry \cite{Fareghbal:2008dy}. The point is that in $4d$, only the non-supersymmetric black holes with a regular horizon and $\mu \ne 0$ can exist, so only solution II is acceptable. The supersymmetric ones with $\mu=0$, as studied in \cite{Leblond:2002ke}, are named superstars and are just naked singularities \cite{Fareghbal:2008dy, Leblond:2002ke}.

The entropy density in the general regular case is
\begin{equation}
s=\frac{\sqrt{(q+r_H)^3(q^\prime+r_H)}}{4G}.
\end{equation}
For the first solution of the extremality condition \eqref{eq:extg}, $q=-r_H$ and $s$ reduces to zero. However, for the second solution for the extremal regular 3+1-charge case, the entropy density, which is the entropy of the ground state, is
\begin{equation}
s_{ext}=\frac{1}{4G}(q+r_H)^2\left(\frac{r_H}{q-2r_H}\right)^{1/2}
=\frac{1}{4G}(q^\prime+r_H)^2 \left(\frac{3r_H}{q^\prime}\right)^{3/2}.
\end{equation}
As we will demonstrate below by considering the order in which limits are taken to go from a regular 3+1-charge black brane to the extremal 3-charged black brane, a discontinuity in various parameters, such as the entropy density, exists. This can be interpreted in the dual $\mathcal{N}=4$ SYM theory as a gap in the states. Within this gap, the system shows the behavior of a Fermi liquid with long-lived quasiparticles and stable fluctuation modes, while outside the gap the quasiparticles are short-lived and the states show non-Fermi-liquid behavior, similar to the $5d$ case in \cite{DeWolfe:2013fha}. 

One should that the new developments in quantum corrections of black holes and perspective from higher dimensions on these effects could also be considered as in \cite{Chen:2025rcc, Nian:2025oei, David:2021qaa}.

\subsection{3-charge black brane}
We first consider the case where $q^\prime=0$ and $q\neq 0$. Then, for the general three-charge black brane solution, we have
\begin{gather}
f(r)=\frac{1}{r L^2}\big((r+q)^3-(r_H+q)^3\big), \quad
\mu=\frac{(q+r_H)^3}{L^2},\nonumber\\
A_t=\left(\frac{q^2}{(q+r_H)^2}+\frac{q(q+r_H)}{L^2}\right)^{1/2}
\left(1-\frac{q+r_H}{q+r}\right), \quad
A_t^\prime=0, \quad
X=\left(1+\frac{q}{r}\right)^{-1/4}, \quad
X^\prime=\left(1+\frac{q}{r}\right)^{3/4},\nonumber\\
T_H=\frac{3}{L^2 \pi}\sqrt{r_H (q+r_H)}, \quad
s=\frac{\sqrt{r_H (q+r_H)^3}}{4G}.
\end{gather}
For the extremal case $T_H=0$, $S=0$, which corresponds to the two solutions of \eqref{eq:extg},
\begin{gather}\label{eq:ext1}
r_H=0 \quad \text{or} \quad q=-r_H \quad \text{(extremal 3-charge black branes)},\nonumber\\
\mu_{\text{ext}}=\frac{4q^3}{L^2} \quad \text{or} \quad 0.
\end{gather}
Now we consider two scenarios for going from the regular non-extremal 3+1-charge black brane to the extremal 3-charge black brane. If we first let $q'\to 0$ and then apply the extremality condition \eqref{eq:ext1}, then
\begin{equation}
\frac{qq'}{r_H^2}\to 0.
\end{equation}
On the other hand, by first applying the extremality condition \eqref{eq:extg} and then letting $q' \to 0$, one gets
\begin{equation}
q=\frac{3r_H^2}{q'}+2r_H, \quad \text{or} \quad \frac{qq'}{r_H^2}\to 3.
\end{equation}
The difference between these two limits indicates the existence of a discontinuity and therefore a gap in the states of the dual field theory. The quantum corrections to this gap then also can be considered, and its behavior from various dimensions can be examined.

\subsection{1-charge black brane}
If we assume $q=0$ and $q'\neq 0$, which is the 1-charge black brane case, then
\begin{gather}
f(r)=\frac{1}{L^2}\left( r(q'+r)-\frac{r_H^2}{r}(q'+r_H)\right), \quad
\mu=\frac{r_H^2(q'+r_H)}{L^2},\nonumber\\
A_t=0, \quad
A_t'=\left( \frac{q'^2}{(q'+r_H)^2}+\frac{q' r_H^2}{(q'+r_H)L^2} \right)^{1/2}
\left(1-\frac{q'+r_H}{q'+r}\right), \quad
X=\left(1+\frac{q'}{r}\right)^{1/4}, \quad
X'=\left(1+\frac{q'}{r}\right)^{-3/4}, \nonumber\\
T_H=\frac{2q'+3r_H}{L^2 \pi}\sqrt{\frac{r_H}{q'+r_H}}, \quad
s=\frac{\sqrt{r_H^3 (q'+r_H)}}{4G}.
\end{gather}
Again, for the extremal case $T_H=0$, $S=0$ and
\begin{gather}\label{eq:ext2}
r_H=0 \quad \text{or} \quad q'=-\frac{3 r_H}{2} \quad \text{(extremal 1-charge black branes)},\nonumber\\
\mu_{\text{ext}}=0 \quad \text{or} \quad -\frac{r_H^3}{2L^2}.
\end{gather}
If we first let $q\to 0$ and then apply the extremality condition, $\frac{qq'}{r_H^2}$ is zero, and if we first use the condition \eqref{eq:extg}, then we have
\begin{equation}
q'=\frac{3 r_H^2}{q-2r_H}, \quad
\frac{qq'}{r_H^2}=\frac{3q}{q-2r_H} \xrightarrow{q\to 0} 0.
\end{equation}
We can also use other ratios to verify the discontinuity in the states arising from the order of the limits. For instance, if we first apply $q\to 0$ and then $r_H\to 0$, we find
\begin{equation}
\frac{q}{q'}\to 0, \quad \frac{r_H}{q'}\to 0, \quad \frac{q}{r_H}\to 0.
\end{equation}
On the other hand, if we first apply \eqref{eq:extg} and then take $q\to 0$, we get
\begin{equation}
\frac{q}{q'}=\frac{q(q-2r_H)}{3r_H^2}\to 0, \quad
\frac{r_H}{q'}=\frac{r_H(q-2r_H)}{3r_H^2}\to -\frac{2}{3}, \quad
\frac{q}{r_H}\to 0.
\end{equation}
The difference between the limits of these parameters again indicates the existence of the gap. This gap  in the states, which arises from this non-commutativity of two limits, is related to the range of energies where there are no stable fermionic quasiparticle excitations.

One should note that inside the gap, the system behaves like Fermi liquid where quasiparticles are long-lived, and their fluctuations are stable, while outside the gap, the system exhibits non-Fermi liquid or strange metal behavior where quasiparticles are short-lived, and the dynamics are dominated by strong interactions with no well-defined particle excitations. So this gap is like a switch, separating stable, conventional physics from the exotic, strongly coupled physics that is a primary focus of holographic duality. This gap is ultimately related to the geometry of the black brane's near-horizon limit.

\begin{figure}[ht!]
\centering
\begin{minipage}{0.9\textwidth}
\centering
\includegraphics[scale=1]{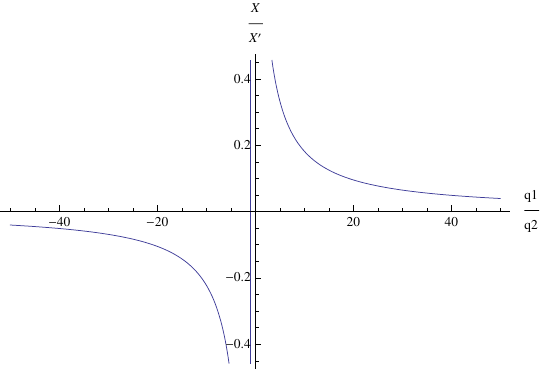}
\caption{The ratio of the dilaton fields $\frac{X}{X'}=\frac{q'+r}{q+r}$ for $r=1$. When $\frac{q}{q'}\to\infty$, $\frac{X}{X'}\to 0$. Also, for fixed $q$ and $q'$, as $r\to\infty$, $\frac{X}{X'}\to 1$. Thus, the dilaton fields are equal at the boundary.}
\label{fig:chem}
\end{minipage}
\end{figure}

\begin{figure}[ht!]
\centering
\begin{minipage}{0.45\textwidth}
\centering
\includegraphics[scale=0.6]{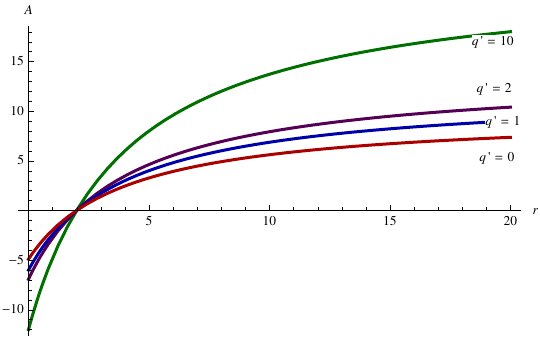}
\caption{Plot of $A$ as a function of $r$ for different $q'$ (with $r_H=2,\ L=1$).}
\label{fig:A111}
\end{minipage}
\hfill
\begin{minipage}{0.45\textwidth}
\centering
\includegraphics[scale=0.6]{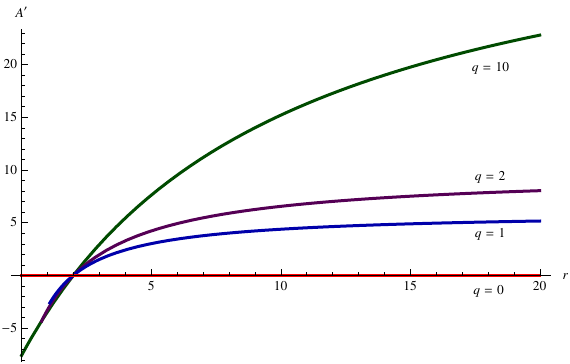}
\caption{Plot of $A'$ as a function of $r$ for different $q$. As $q\to 0$, the curves approach the $x$-axis. At $q=0$, $A'$ vanishes identically (with $r_H=2,\ L=1$).}
\label{fig:Appp2}
\end{minipage}
\end{figure}

Based on \cite{Fareghbal:2008dt}, the near-horizon of these charged black holes can "decouple" from the rest of the spacetime. This decoupled sector is described by a 2D CFT on the gravity side, which is dual to a specific, large R-charge sector of the original $3d$ field theory. Therefore, this fragile gap can be seen as the field theory signature of the system entering this stable, decoupled, lower-dimensional regime. The moment $q'$ is turned on, even by a tiny amount, the gap vanishes and this is related to a discontinuity in the corresponding chemical potential.

This extreme sensitivity suggests the gap is a critical phenomenon, a phase transition point that exists only at a specific, fine-tuned value of a parameter $(q'=0)$. It is a sharp boundary in the theory's parameter space.

The singular near-horizon geometry of the $4d$ black brane, which is responsible for the gap, becomes a smooth and non-singular $\text{AdS}_3 \times \mathbb{R}^3$ geometry when uplifted to six dimensions. It is a region where the quasiparticle's momentum becomes spacelike in this six-dimensional spacetime. This happens because the momentum component along the extra Kaluza-Klein dimension, which is associated with charge $q'$, is crucial. This also explains the $q'$ dependence, as when $q'=0$, the momentum component vanishes, the quasiparticle momentum becomes spacelike, and a stable gap opens up.

So this fragile gap is a deeply geometric signature of a quantum critical point, which reveals a hidden, lower dimensional structure within a strongly coupled system, as this gap acts like a ``stable island," isolating conventional Fermi liquid behavior from the surrounding non-Fermi liquid background. These decoupled degrees of freedom also have their own separate Hilbert space and do not mix with the rest of the theory.

Again, one should note that the gap is not a property of the extremal 3-charge black hole itself, but rather a memory of the order in which the 3+1-charge system was taken to the 3-charge limit. So it suggests that in the dual field theory, the state of the system depends on the history of how the parameters were tuned, reminiscent of hysteresis or the emergence of non-equilibrium phases in strongly coupled systems. Also, this behavior and this critical point is universal and has scale invariance, so we expect to see it in various dimensions as well.

\subsection{Chemical potentials}

The chemical potentials of these black hole solutions are

\begin{gather}
\mu_1=\frac{\sqrt{3}}{L} \left(\frac{q^2}{(q+r_H)^2}+\frac{q(q'+r_H)(q+r_H)}{L^2 r_H}\right), \ \ \ \ \ 
\mu_2=\frac{\sqrt{3}}{L} \left(\frac{q'^2}{(q'+r_H)^2}+\frac{q'(q+r_H)^3}{(q'+r_H) L^2 r_H}\right).
\end{gather}

The factor of $\frac{\sqrt{3}}{L}$ is for canonically normalizing the chemical potentials. From these, it can be seen that if we first let $q'=0$ and then $r_H \to 0$, the limit of $\mu_1$ is $\mu_1 \to \frac{\sqrt{3} q^2}{L^3 r_H}$. However, if we first go to the extremal limit by letting $r_H \to 0$ and then $q' \to 0$, then $\mu_1 \to 0$. This behavior of the chemical potential, which is associated with the boundary value of the gauge field, again indicates the gap in the CFT. For $\mu_2$, by changing $q'$ from a positive to a negative value, the chemical potential jumps from a positive finite value to a negative finite value. This discontinuity indicates that some states are missing in between, and therefore a gap should be present on the field theory side.

\begin{figure}[ht!]
\centering
\begin{minipage}{0.45\textwidth}
\centering
\includegraphics[scale=1]{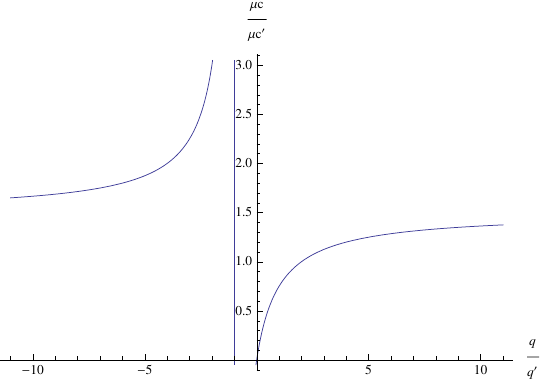}
\caption{$\frac{\mu_c}{\mu_c'}$ as a function of $\frac{q}{q'}$ for $r_H=1$, $L=1$, $q'=2$.}
\label{fig:chemp}
\end{minipage}
\hfill
\begin{minipage}{0.45\textwidth}
\centering
\includegraphics[scale=1]{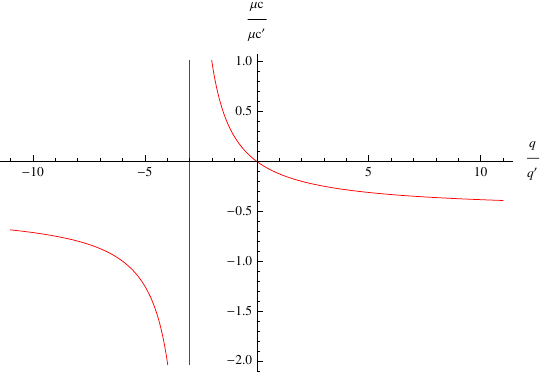}
\caption{$\frac{\mu_c}{\mu_c'}$ as a function of $\frac{q}{q'}$ for $r_H=3$, $L=1$, $q'=-2$.}
\label{fig:chemn}
\end{minipage}
\end{figure}

\begin{figure}[ht!]
\centering
\includegraphics[scale=1]{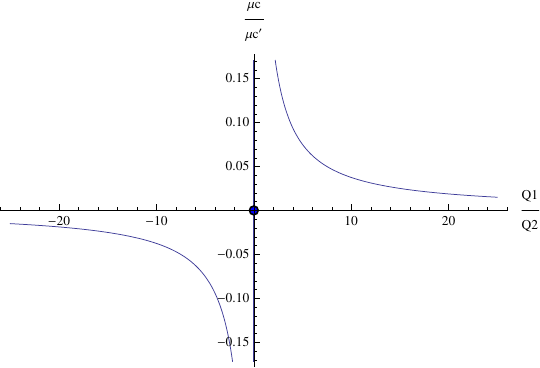}
\caption{The ratio of the chemical potentials $\frac{\mu_c}{\mu_c'}$ as a function of $\frac{Q}{Q'}$, with $r_H = L = 1$ and $Q = 3$.}
\label{fig:chemQ}
\end{figure}

As the chemical potential around $q'=0$ does not change continuously, and there is a jump around the center, as can be seen from figures \ref{fig:chemp}, \ref{fig:chemn} and \ref{fig:chemQ}, the density of states does not change continuously everywhere, and a gap could appear. Within this gap, the chemical potential remains constant and is different from the values of the chemical potential around it. Therefore, the degrees of freedom inside and outside the gap should be decoupled. \\

In the right panel, Figure \ref{fig:chemn}, as $\frac{\mu_c}{\mu_c'}$ decreases with increasing $\frac{q}{q'}$, the susceptibility is negative. Due to the perturbative instability in the spectrum of the bosonic fluctuations, the thermodynamic system is unstable.

In equilibrium, the sum of the chemical potentials is zero, as the free energy is at its minimum. During phase changes or during any reaction, the chemical potential changes from higher values to lower values, and free energy is released. From these Figures, \ref{fig:chemp}, \ref{fig:chemn} and \ref{fig:chemQ}, when $\frac{q}{q'}$ increases—for example, if the number of charge 1 relative to charge 2 is increasing—the ratio of the chemical potentials $\frac{\mu_1}{\mu_2}$ decreases, which means the system is moving further away from equilibrium and is therefore thermodynamically unstable. This is due to the particle creation effects around this gap.

Since the gap vanishes when $q'$ is turned on, because of the discontinuity in the chemical potential $\mu_1$, it is a holographic realization of a Lifshitz transition. This is a change in the topology of the Fermi surface, where the chemical potential jumps discontinuously.

\section{Field theory duals}
Our aim in this section is to holographically study a three-dimensional $\mathcal{N}=4$ SYM CFT using an extremal charged black brane solution and considering all the fermionic modes.

The gauge group here is $\text{SU}(N)$ with a large number of colors and a large 't Hooft coupling at zero temperature, and with four non-zero chemical potentials, three of which are equal with our choice of gauge fields.

Now we gauge $\text{SO}(8)$ to a Cartan subgroup of it, which is a $\mathcal{N}=2$ theory with the subgroup $U(1)_a \times U(1)_b \times U(1)_c \times U(1)_d$. The charges of the dual gauginos (adjoint Majorana fermions) in the $\mathbf{4}$ of $\text{SO}(8)$, and the dual scalars $Z_j = X_{2j-1} + iX_{2j}$, $j = 1,2,3,4$, in the $\mathbf{8}$ of $\text{SO}(8)$, are given in the table below, where $q_1 = q_a$, $q_2 = q_b + q_c + q_d$, and $q_3 = q_1 + q_2$.

\begin{center}
    \begin{tabular}{| c | c | c | c | c | c | c | c | c |}
    \hline
          & $\lambda_1$ & $\lambda_2$ & $\lambda_3$ & $\lambda_4$ & $Z_1$ & $Z_2$ & $Z_3$ & $Z_4$ \\ \hline
    $q_a$ & $\frac{1}{2}$ & $\frac{1}{2}$ & $-\frac{1}{2}$ & $-\frac{1}{2}$ & $1$ & $0$ & $0$ & $0$ \\ \hline
    $q_b$ & $\frac{1}{2}$ & $-\frac{1}{2}$ & $\frac{1}{2}$ & $-\frac{1}{2}$ & $0$ & $1$ & $0$ & $0$ \\ \hline
    $q_c$ & $\frac{1}{2}$ & $-\frac{1}{2}$ & $-\frac{1}{2}$ & $\frac{1}{2}$ & $0$ & $0$ & $1$ & $0$ \\ \hline
    $q_d$ & $\frac{1}{2}$ & $-\frac{1}{2}$ & $-\frac{1}{2}$ & $-\frac{1}{2}$ & $0$ & $0$ & $0$ & $1$ \\ \hline
    $q_1$ & $\frac{1}{2}$ & $\frac{1}{2}$ & $-\frac{1}{2}$ & $-\frac{1}{2}$ & $1$ & $0$ & $0$ & $0$ \\ \hline
    $q_2$ & $\frac{3}{2}$ & $-\frac{3}{2}$ & $-\frac{1}{2}$ & $-\frac{1}{2}$ & $0$ & $1$ & $1$ & $1$ \\ \hline
    $q_3$ & $2$ & $-1$ & $-1$ & $-1$ & $1$ & $1$ & $1$ & $1$ \\ \hline
    \end{tabular}
\end{center}

The vacuum expectation value for the dual operator on the boundary of the scalar field $\phi(r)$ in the $U(1)^4$ gauged supergravity background is
\begin{equation}
\mathcal{O}_{\mathbf{20^\prime}} \sim \operatorname{Tr}\left(-3|Z_1|^2 + |Z_2|^2 + |Z_3|^2 + |Z_4|^2\right).
\end{equation}
The sign of this value changes at the extremal 3+1-charge black holes where
\begin{equation}
\mu_R = \left(\frac{q}{q'}\right)^{3/2} \frac{2q' + \sqrt{q'^2 + 3q q'}}{(3q - q') + \sqrt{q'^2 + 3q q'}}.
\end{equation}

Now we classify the 56 fermion modes by their charges using this Cartan subgroup. The gravitini only can get a charge of $\frac{1}{2}$ or $-\frac{1}{2}$. So the modes which have only these charges can mix with the gravitini. Therefore the 32 modes which are mixing with the gravitini would be

\begin{center}
    \begin{tabular}{ | l | l | l | p{5cm} |}
    \hline
    Fermion Modes & $q_1$ & $q_b+q_c+q_d$ & Dual Operators \\ \hline
    ${\chi_1}^{(\frac{1}{2},\frac{1}{2}, \frac{1}{2},\frac{1}{2})}$, ${\chi_2}^{(\frac{1}{2},\frac{1}{2},\frac{1}{2},\frac{1}{2})}$, ${\chi_3}^{(\frac{1}{2}, \frac{1}{2},\frac{1}{2},\frac{1}{2})}$,${\chi_4}^{(\frac{1}{2},\frac{1}{2},\frac{1}{2},\frac{1}{2})}$ & $\frac{1}{2}$ & $\frac{3}{2} $& $\lambda_1 Z_1, \lambda_1 Z_2 , \lambda_1 Z_3 , \lambda_1 Z_4$  \\ \hline
    ${\chi_1}^{(\frac{1}{2},-\frac{1}{2}, -\frac{1}{2},-\frac{1}{2})}$,${\chi_2}^{(\frac{1}{2},-\frac{1}{2},-\frac{1}{2},-\frac{1}{2})}$,${\chi_3}^{(\frac{1}{2}, -\frac{1}{2},-\frac{1}{2},-\frac{1}{2})}$,${\chi_4}^{(\frac{1}{2},-\frac{1}{2},-\frac{1}{2},\frac{-1}{2})}$ & $\frac{1}{2}$ & $-\frac{3}{2}$ & $ \lambda_2 Z_1,  \lambda_2 Z_2,  \lambda_2 Z_3,  \lambda_2 Z_4$ \\ \hline
    ${\chi_1}^{(-\frac{1}{2},-\frac{1}{2},0,0)}, {\chi_2}^{(-\frac{1}{2},-\frac{1}{2},0,0)}, {\chi_3}^{(-\frac{1}{2},-\frac{1}{2},0,0)}, {\chi_4}^{(-\frac{1}{2},-\frac{1}{2},0,0)}$ & $-\frac{1}{2}$ & $-\frac{1}{2}$ & $ \lambda_3 Z_1, \lambda_3 Z_2, \lambda_3 Z_3, \lambda_3 Z_4$ \\ \hline
    ${\chi_1}^{(-\frac{1}{2},0,-\frac{1}{2},0)}, {\chi_2}^{(-\frac{1}{2},0,-\frac{1}{2},0)}, {\chi_3}^{(-\frac{1}{2},0,-\frac{1}{2},0)}, {\chi_4}^{(-\frac{1}{2},0,-\frac{1}{2},0)}$ & $-\frac{1}{2}$ & $-\frac{1}{2}$ & $ \lambda_3 Z_1, \lambda_3 Z_2, \lambda_3 Z_3, \lambda_3 Z_4$ \\ \hline
    ${\chi_1}^{(-\frac{1}{2},0,0,-\frac{1}{2})}, {\chi_2}^{(-\frac{1}{2},0,0,-\frac{1}{2})}, {\chi_3}^{(-\frac{1}{2},0,0,-\frac{1}{2})},{\chi_4}^{(-\frac{1}{2},0,0,-\frac{1}{2})}$ & $-\frac{1}{2}$ & $-\frac{1}{2}$ & $ \lambda_3 Z_1, \lambda_3 Z_2, \lambda_3 Z_3, \lambda_3 Z_4$ \\ \hline
 ${\chi_1}^{(-\frac{1}{2},-\frac{1}{2},0,0)}, {\chi_2}^{(-\frac{1}{2},-\frac{1}{2},0,0)}, {\chi_3}^{(-\frac{1}{2},-\frac{1}{2},0,0)}, {\chi_4}^{(-\frac{1}{2},-\frac{1}{2},0,0)}$ & $-\frac{1}{2}$ & $-\frac{1}{2}$ & $ \lambda_4 Z_1, \lambda_4 Z_2, \lambda_4 Z_3, \lambda_4 Z_4$ \\ \hline
    ${\chi_1}^{(-\frac{1}{2},0,-\frac{1}{2},0)}, {\chi_2}^{(-\frac{1}{2},0,-\frac{1}{2},0)}, {\chi_3}^{(-\frac{1}{2},0,-\frac{1}{2},0)}, {\chi_4}^{(-\frac{1}{2},0,-\frac{1}{2},0)}$ & $-\frac{1}{2}$ & $-\frac{1}{2}$ & $ \lambda_4 Z_1, \lambda_4 Z_2, \lambda_4 Z_3, \lambda_4 Z_4$ \\ \hline
    ${\chi_1}^{(-\frac{1}{2},0,0,-\frac{1}{2})}, {\chi_2}^{(-\frac{1}{2},0,0,-\frac{1}{2})}, {\chi_3}^{(-\frac{1}{2},0,0,-\frac{1}{2})},{\chi_4}^{(-\frac{1}{2},0,0,-\frac{1}{2})}$ & $-\frac{1}{2}$ & $-\frac{1}{2}$ & $ \lambda_4 Z_1, \lambda_4 Z_2, \lambda_4 Z_3, \lambda_4 Z_4$ \\ \hline
\end{tabular}
\end{center}

The 32 modes that do not mix with the gravitini, because they have a maximum charge of 2 that the gravitini cannot carry, are

\begin{center}
    \begin{tabular}{ | l | l | l | p{5cm} |}
    \hline
    Fermion Modes & $q_1$ & $q_b+q_c+q_d$ & Dual Operators \\ \hline
    ${\chi}^{(2,\frac{1}{2},\frac{1}{2},\frac{1}{2})}$ & $2$ & $\frac{7}{2} $& $\lambda_1 Z_1$  \\ \hline
    ${\chi}^{(2,-\frac{1}{2},-\frac{1}{2},-\frac{1}{2})}$ & $2$ & $\frac{1}{2}$ & $ \lambda_2 Z_1$ \\ \hline
    ${\chi}^{(2,-\frac{1}{2},0,0)}, {\chi}^{(2,0,-\frac{1}{2},0)}, {\chi}^{(2,0,0,-\frac{1}{2})}$ & $2$ & $\frac{3}{2}$ & $\lambda_3 Z_1 ,\lambda_3 Z_1,\lambda_3 Z_1$ \\ \hline
    ${\chi}^{(2,-\frac{1}{2},0,0)}, {\chi}^{(2,0,-\frac{1}{2},0)}, {\chi}^{(2,0,0,-\frac{1}{2})}$ & $2$ & $\frac{3}{2}$ & $ \lambda_4 Z_1,\lambda_4 Z_1,\lambda_4 Z_1$ \\ \hline
   ${\chi}^{(\frac{1}{2},2,\frac{1}{2},\frac{1}{2})}, {\chi}^{(\frac{1}{2},\frac{1}{2},2,\frac{1}{2})}, {\chi}^{(\frac{1}{2},\frac{1}{2},\frac{1}{2},2)}$ & $\frac{1}{2}$ & $\frac{7}{2}$ & $ \lambda_1 Z_2,  \lambda_1 Z_3 ,  \lambda_1 Z_4$ \\ \hline
    ${\chi}^{(-\frac{1}{2},2,-\frac{1}{2},-\frac{1}{2})}, {\chi}^{(-\frac{1}{2},-\frac{1}{2},2,-\frac{1}{2})}, {\chi}^{(-\frac{1}{2},-\frac{1}{2},-\frac{1}{2},2)}$ & $-\frac{1}{2}$ & $\frac{1}{2}$ & $ \lambda_2 Z_2,  \lambda_2 Z_3 ,  \lambda_2 Z_4$ \\ \hline
    ${\chi}^{(-\frac{1}{2},2,0,0)}, {\chi}^{(-\frac{1}{2},0,2,0)}, {\chi}^{(-\frac{1}{2},0,0,2)}$ & $-\frac{1}{2}$ & $\frac{3}{2}$ & $ \lambda_3 Z_2,  \lambda_3 Z_2 ,  \lambda_3 Z_2$ \\ \hline
    ${\chi}^{(-\frac{1}{2},0,2,0)}, {\chi}^{(0,-\frac{1}{2},2,0)}, {\chi}^{(0,0,2,-\frac{1}{2})}$ & $-\frac{1}{2}$ & $\frac{3}{2}$ & $ \lambda_3 Z_3,  \lambda_3 Z_3 ,  \lambda_3 Z_3$ \\ \hline
    ${\chi}^{(-\frac{1}{2},0,0,2)}, {\chi}^{(0,-\frac{1}{2},0,2)}, {\chi}^{(0,0,-\frac{1}{2},2)}$ & $-\frac{1}{2}$ & $\frac{3}{2}$ & $ \lambda_3 Z_4,  \lambda_3 Z_4 ,  \lambda_3 Z_4$ \\ \hline
    ${\chi}^{(-\frac{1}{2},2,0,0)}, {\chi}^{(0,2,-\frac{1}{2},0)}, {\chi}^{(0,2,0,-\frac{1}{2})}$ & $-\frac{1}{2}$ & $\frac{3}{2}$ & $ \lambda_4 Z_3,  \lambda_4 Z_3 ,  \lambda_4 Z_3$ \\ \hline
    ${\chi}^{(-\frac{1}{2},0,0,2)}, {\chi}^{(0,-\frac{1}{2},0,2)}, {\chi}^{(0,0,-\frac{1}{2},2)}$ & $-\frac{1}{2}$ & $\frac{3}{2}$ & $ \lambda_4 Z_4,  \lambda_4 Z_4 ,  \lambda_4 Z_4$ \\ \hline
    \end{tabular}
\end{center}

We will only be interested in these modes, and the Dirac equation will only be solved for them.\\
As can be noted, the modes with dual operators consisting of $\lambda_3$ or $\lambda_4$ are degenerate. 
Note that this degeneracy is due to the fact that the $SO(8)$ R-symmetry of the $N=8$ supergravity is being broken down to $U(1)^4$, and some modes become equivalent under this broken symmetry.

One should note that $64 = 56 \oplus 8$, so 8 modes from the above charts are being decomposed and eliminated from the rest. One could also find quantum corrections of these modes and therefore finds the quantum corrections of the fermionic responce and the gap.

\section{Charged black holes from $\mathcal{N}=8$ maximal gauged supergravity}
In order to derive the Dirac equation for the fluctuations of fermions, we need to embed the Lagrangian \ref{eq:lagmain} in the full maximal $\mathcal{N}=8$ four-dimensional gauged supergravity theory.

The supermultiplet of $\mathcal{N}=8$ supergravity consists of a spin-2 graviton $e_\mu^a$, 8 spin-$\frac{3}{2}$ gravitini $\psi_\mu^i$, 28 spin-1 fields $A_\mu^{IJ}$, 56 spin-$\frac{1}{2}$ fermions $\chi^{ijk}$, and 70 spin-0 (35 complex scalar fields) states. Gravitons always behave as a singlet under any internal symmetry group, and fermions sit in the $\mathbf{8}$- and $\mathbf{56}$-dimensional representations of chiral $SU(8)$. However, the largest internal symmetry group that the $\mathbf{28}$ vector fields allow is $SO(8)$, as they cannot transform under a complex internal symmetry, but this can be extended to $E_{7(+7)}$ by the generalized duality transformations \cite{DeWit:1982td}. The scalar fields parameterize an $E_7 / SU(8)$ coset manifold, into which the $SO(8)$ gauge group can be embedded.

The $E_{7(7)}$ elements $u$ and the scalar coset representatives $v$ sit in the $56 \times 56$ matrix $\mathcal{V}$,
\[
\mathcal{V} =
\begin{pmatrix}
u_{ij}{}^{IJ} & v_{ijKL} \\
v^{klIJ} & u^{kl}{}_{KL}
\end{pmatrix}.
\]
The covariant derivative for the 56 spin-$\frac{1}{2}$ Majorana fermions $\chi_{ijk} = \chi_{[ijk]}$, with $i,j,k = 1,\dots,8$, is
\begin{equation}
D_\mu \chi_{ijk} = \nabla_\mu \chi_{ijk} + 3g A_\mu^{\ m}{}_{[i} \chi_{jk]m} + 3g A_\mu'^{\ m}{}_{[i} \chi_{jk]m},
\end{equation}
where $g = \frac{1}{2L}$.

The full Lagrangian of $\mathcal{N}=8$ supergravity \cite{DeWit:1982td}, describing all the matter fields of the theory with any possible coupling, is
\begin{gather}
\mathcal{L}=-\frac{1}{2}e R(e,\omega)-\frac{1}{2} \epsilon^{\mu\nu\rho\sigma}(\bar{\psi}^i_\mu \gamma_\nu D_\rho \psi_{\sigma i}-\bar{\psi^i_\mu} \overleftarrow{D} \gamma_\nu \psi_{\sigma i})-\frac{1}{12} e(\bar{\chi}^{ijk}\gamma^\mu D_\mu \chi_{ijk}-\bar{\chi^{ijk}} \overleftarrow{D}_\mu \gamma^\mu \chi_{ijk})\nonumber\\
-\frac{1}{96} e {\mathcal{A}_\mu}^{ijkl} {\mathcal{A}^\mu}_{ijkl}-\frac{1}{8}[F^{+}_{\mu\nu IJ} (2S^{IJ,KL}-{\delta^{IJ}}_{KL}) {F^{+\mu\nu}}_{KL}+h.c.]-\frac{1}{2}e[F^{+\mu\nu IJ} S^{IJ,KL} O^{+\mu\nu KL}+h.c.]\nonumber\\
-\frac{1}{4} e [{O^{+IJ}}_{\mu\nu} (S^{IJ,KL}+{u^{ij}}_{IJ} v_{ijKL}) O^{+\mu\nu KL}+h.c.]-\frac{1}{24}e[\bar{\chi}_{ijk} \gamma^\nu \gamma^\mu \psi_{\nu l}({{\mathcal{A}}_\mu}^{ijkl}+{\mathcal{A}_\mu}^{ijkl})+h.c.]
-\frac{1}{2} e {{\bar{\psi}}^{[i}}_\mu  \psi^{j]}_\nu {\bar{\psi}^\mu}_i {\psi^\nu}_j  \nonumber\\ +\frac{\sqrt{2}}{4}e[{\bar{\psi}^i}_\lambda \sigma^{\mu\nu}\gamma^\lambda \chi_{ijk} {\bar{\psi}_\mu}^j {\psi^k}_\nu+h.c.] +e[\frac{1}{144}\eta \epsilon_{ijklmnpq} {\bar{\chi}}^{ijk}\sigma^{\mu\nu} \chi^{lmn}{\bar{\psi}^p}_\mu {\psi^{q}}_\nu  +\frac{1}{8}\ { \bar{\psi}^i}_\lambda \sigma^{\mu\nu} \gamma^\lambda \chi_{ikl} { \bar{\psi}_{\mu j}}\gamma_\nu \chi^{jkl} +h.c.] \nonumber\\
+\frac{\sqrt{2} }{864} \eta e [\epsilon^{ijklmnpq} {\bar{\chi}}_{ijk} \sigma^{\mu\nu} \chi_{lmn} {\bar{\psi}_\mu}^r \gamma_\nu \chi_{pqr}+h.c.]+\frac{1}{32} e \bar{\chi}^{ikl} \gamma^\mu \chi_{jkl} \bar{\chi}^{jmn}\gamma_\mu \chi_{imn}-\frac{1}{96} e (\bar{\chi}^{ijk} \gamma^\mu \chi_{ijk})^2.
\end{gather}

We can truncate this Lagrangian to the quadratic terms that describe fermion dynamics. We are also interested in terms that do not couple fermions to the gravitini \cite{DeWolfe:2011ts}. So the fermionic Lagrangian is
\begin{gather}
\mathcal{L}_{1/2}=-\frac{1}{12}e {\bar{\chi}}^{ijk}(\gamma^\mu D_\mu-\overleftarrow{D}_\mu \gamma^\mu)\chi_{ijk}-\frac{1}{2}  e ({F^{+}}_{\mu\nu i j} S^{ij,kl} O^{+\mu\nu k l} +h.c.).
\end{gather}
Here $O^{+\mu\nu i j}$ is
\begin{gather}
O^{+\mu\nu i j }=-\frac{\sqrt{2}}{144} \eta \epsilon^{ijklmnpq} {\bar{\chi}}_{klm} \sigma_{\mu\nu} \chi_{npq}-\frac{1}{2} {\bar{\psi}}_{\lambda k}\sigma_{\mu\nu} \gamma^\lambda \chi^{ijk}+\frac{\sqrt{2}}{2} {\bar{\psi}^i}_\rho \gamma^{[\rho} \sigma_{\mu\nu} \gamma^{\sigma ]} {\psi^j}_\sigma.
\end{gather}
$\eta$ can take $\pm 1$.

The $E_{7(7)}$ group can be decomposed as \cite{Ceresole:2009jc}
\begin{align}
E_{7(7)}  \ \ \ \ \ \ \ \ \ \ \ \ \  \longrightarrow \ \ \ \ \ \ \ \ \ \ \ \ \ \ \ \ \ \ \ \ \ \ \ \ \ \ \ \ \ \ \ \ \ \ \ \ \ \ SL(8, \mathbb{R} ) \nonumber\\
\downarrow  \ \ \ \ \ \ \ \ \ \  \ \ \ \ \ \ \ \ \ \ \ \ \ \ \ \ \ \ \ \ \ \ \ \ \ \ \ \ \ \ \ \ \ \ \ \ \ \ \ \ \ \ \ \ \ \ \ \  \ \ \ \ \ \ \ \  \downarrow \ \ \ \nonumber\\
E_{6(6)} \times SO(1,1)\ \ \ \ \ \ \ \ \ \ \longrightarrow  \ \ \ \ \ \  \ \ SL(6, \mathbb{R})\times SL(2,\mathbb{R}) \times SO(1,1).
\end{align}
\begin{gather}
E_{7(7)} \to E_{6(6)} \times SO(1,1), \nonumber\\
E_{6(6)} \to SL(6,\mathbb{R}) \times SL(2,\mathbb{R}),\nonumber\\
\mathbf{56} \to 28+28^\prime \to \Big \{ (15,1,1)+(6^\prime,2,1)+(1,1,3)+ (15^\prime,1,-1)+(6,2,-1)+(1,1,-3)\Big\},\ \  \text{or} \nonumber\\
\mathbf{56} \to (\mathbf{27},1)+(\mathbf{1},3 )+({\mathbf{27}}^\prime, -1)+({\mathbf{1}}^\prime,-3) \to \nonumber\\ \Big \{ (15,1,1)+(6^\prime,2,1)+(1,1,3)+ (15^\prime,1,-1)+(6,2,-1)+(1,1,-3)\Big\}. \nonumber\\
\end{gather}

Here, $\mathbf{27}$ is in the real part of $E_{6(6)}$, and $E_{6(6)} \times SO(1,1)$ is a maximal, non-compact, and symmetric embedding in $E_{7(7)}$. So if we move horizontally in that diagram, $E_{7(7)}$ decomposes symmetrically into $\mathbf{28} + \mathbf{28^\prime}$.

In these equations, $S^{ij,kl}$ is a function of the scalars. The scalars for the 3+1-charge black hole live in $SL(8, \mathbb{R})$,
\begin{equation}
S = \operatorname{diag} \left\{ e^{\frac{\sqrt{3}\phi}{6}}, e^{\frac{\sqrt{3}\phi}{6}}, e^{-\frac{\phi}{6\sqrt{3}}}, e^{-\frac{\phi}{6\sqrt{3}}}, e^{-\frac{\phi}{6\sqrt{3}}}, e^{-\frac{\phi}{6\sqrt{3}}}, e^{-\frac{\phi}{6\sqrt{3}}}, e^{-\frac{\phi}{6\sqrt{3}}} \right\}.
\end{equation}
We can decompose the $SL(8,\mathbb{R})$ indices $I,J = 1,\dots,8$ into $x,y = 1,2$ and $i,j = 3,\dots,8$. The $SL(2,\mathbb{R})$ element is $S' = \mathbb{I}$, so
\begin{equation}
S^y{}_x = e^{\frac{\sqrt{3}\phi}{6}} \delta^y{}_x, \qquad
S^j{}_i = e^{-\frac{\phi}{6\sqrt{3}}} \delta^j{}_i, \qquad
S^x{}_i = S^i{}_x = 0, \qquad
(S')^\alpha{}_\beta = \delta^\alpha{}_\beta.
\end{equation}
Then one can find the elements of $u$ and $v$ using the relations in \cite{DeWit:1982td},
\begin{align}
(u^{ij}{}_{IJ} + v^{ijIJ}) S^{IJ,KL} &= u^{ij}{}_{KL}, \\
(u^{-1})^{IJ}{}_{ij} \, u^{ij}{}_{KL} &= \delta^{IJ}{}_{KL}.
\end{align}

\subsection{Scalar potential}

The general form of the four-dimensional scalar field potential is 
\begin{gather}
V(\phi)=\frac{g^2}{24} {A^i}_{2jkl} \big ({A^i}_{2jkl} \big )^* -\frac{3 g^2}{4}  {A^{ij}}_1 \big ({A^{ij}}_1 \big )^ *.
\end{gather} 

The matrices $A$ for the four-dimensional $\mathcal{N}=8$ supergravity theory are
\begin{gather}
{A_1}^{ij}=-\frac{4}{21} {T_m}^{ijm},\ \ \ \ \ \ \ \
{A_{2 \ell}}^{ijk}=-\frac{4}{3} {T_\ell}^{i^\prime j^\prime k^\prime} \delta _{i^\prime j^\prime k^\prime }^{ijk}, \nonumber\\
A_{3ijk,lmn}=-\frac{\sqrt{2}}{108} \eta \epsilon_{ijkpqr[lm} {T_{n]}}^{par}.
\end{gather}

The $SU(8)$ $T$-tensor coming from the local gauge coupling is
\begin{equation}
T_\ell^{kij} \equiv \left(u^{ij}{}_{IJ} + v^{ijIJ}\right) \left(u_{\ell m}{}^{JK} u^{km}{}_{KI} - v_{\ell mJK} v^{kmKI}\right).
\end{equation}

The elements of 56-bein of $\mathcal{V}$ are
\begin{gather}
{{\mathcal{V}}^{\mathcal{\tilde{A}}}}_{\tilde{\mathcal{B}}}=\text{exp} {\Big(\sum_n \phi_n g^{(n)} \Big)^{\tilde{\mathcal{A}}}}_{\tilde{\mathcal{B}}}.
\end{gather}
Also from \cite{Fischbacher:2009cj}, we can read
\begin{eqnarray}
{u_{ij}}^{IJ}= 2 {\mathcal{V}^{\tilde{\mathcal{A}}}}_{\tilde{\mathcal{B}}} \  \delta^{m}_{\tilde{\mathcal{A}}} \ \delta^{\tilde{\mathcal{B}}}_{n} \  \delta^{ab}_{ij}\ \delta^{IJ}_{cd} \ \ \
\text{for} \ \ \ \tilde{\mathcal{A}} \le 28, \ \tilde{\mathcal{B}} \le 28, (a,b)= Z(m), (c,d)=Z(n), \nonumber\\ \nonumber\\
{u^{kl}}_{KL}=  2 {\mathcal{V}^{\tilde{\mathcal{A}}}}_{\tilde{\mathcal{B}}} \ \delta^{m}_{\tilde{\mathcal{A}}} \ \delta^{\tilde{\mathcal{B}}}_{n} \  \delta^{kl}_{ab} \ \delta^{cd}_{KL}  \ \ \
\text{for} \ \ \ \tilde{\mathcal{A}} \ge 28, \ \tilde{\mathcal{B}} \ge 28, (a,b)= Z(m-28), (c,d)=Z(n-28) ,\nonumber\\ \nonumber\\
v^{klIJ}= 2 {\mathcal{V}^{\tilde{\mathcal{A}}}}_{\tilde{\mathcal{B}}} \ \delta^{m}_{\tilde{\mathcal{A}}} \ \delta^{\tilde{\mathcal{B}}}_{n} \ \delta^{ab}_{ij} \ \delta^{cd}_{KL} \ \ \
\text{for} \ \ \ \tilde{\mathcal{A}} \le 28, \ \tilde{\mathcal{B}} > 28, (a,b)= Z(m), (c,d)=Z(n-28) , \nonumber\\ \nonumber\\
{v^{klIJ}}=  2 {\mathcal{V}^{\tilde{\mathcal{A}}}}_{\tilde{\mathcal{B}}} \ \delta^{m}_{\tilde{\mathcal{A}}} \ \delta^{\tilde{\mathcal{B}}}_{n} \ \delta^{kl}_{ab} \ \delta^{cd}_{KL}  \ \ \
\text{for} \ \ \ \tilde{\mathcal{A}} \ge 28, \ \tilde{\mathcal{B}} \le 28, (a,b)= Z(m-28), (c,d)=Z(n), \nonumber\\
\end{eqnarray}
where

\[ \delta^{a_1 a_2 ... a_n}_{b_1 b_2 . . . b_n} = \left\{
  \begin{array}{l l l}
  +\frac{1}{n!}  & \quad \text{for $b_1 b_2 . . . b_n$ an even permutation of $a_1 a_2 . . . a_n$,}\\
   -\frac{1}{n!} & \quad \text{for $b_1 b_2 . . . b_n$ an odd permutation of $a_1 a_2 . . . a_n$,}\\
   0 & \quad \text{else.}
  \end{array} \right.\]

This scalar potential has a local maximum at $\phi_n = 0$, where $V = -6g^2 = -\frac{3}{2L^2}$, and there are six other stationary points. Using the mentioned equation, one can find that the scalar potential of the 3+1-charge black holes embedded in $\mathcal{N}=8$ supergravity theory is
\begin{equation}
V(\phi) = -\frac{3}{4L^2} \left( e^{-\frac{\phi}{\sqrt{3}}} + e^{\frac{\phi}{\sqrt{3}}} \right).
\end{equation}

For the $\mathcal{N}=2$ truncation of gauged $\mathcal{N}=8$ supergravity without axions, the potential and superpotential are \cite{Batrachenko:2005fn}
\begin{gather}
V = -g^2 \sum_{i<j} X_i X_j, \qquad W = \frac{1}{2} g \sum_i X_i,
\end{gather}
where $g = \frac{1}{2L}$.

Then, using $X_1 = X_2 = X_3 = e^{-\frac{\phi}{2\sqrt{3}}}$ and $X_4 = e^{\frac{\sqrt{3}\phi}{2}}$, we can find
\begin{gather}
V = -\frac{3}{4L^2} \left( e^{-\frac{\phi}{\sqrt{3}}} + e^{\frac{\phi}{\sqrt{3}}} \right), \qquad
W = \frac{1}{4L} \left( 3 e^{-\frac{\phi}{2\sqrt{3}}} + e^{\frac{\sqrt{3}\phi}{2}} \right).
\end{gather}

We can use some general results that apply to theories built out of chiral superfields. The fermion mass matrix can be found as
\begin{equation}
m_F^{ij} = \frac{\partial^2 W}{\partial \phi_i \partial \phi_j}.
\end{equation}

From this, one can find
\begin{align}
m^{11} &= m^{22} = m^{33} = \frac{1}{16L} \left( 3 e^{-\frac{\phi}{2\sqrt{3}}} + e^{\frac{\sqrt{3}\phi}{2}} \right), \\
m^{13} &= m^{23} = -m^{12} = \frac{1}{16L} \left( e^{-\frac{\phi}{2\sqrt{3}}} - e^{\frac{\sqrt{3}\phi}{2}} \right),
\end{align}
with eigenvalues
\begin{equation}
\frac{1}{4L} e^{-\frac{\phi}{2\sqrt{3}}}, \qquad
\frac{1}{4L} e^{-\frac{\phi}{2\sqrt{3}}}, \qquad
\frac{1}{16L} \left( e^{-\frac{\phi}{2\sqrt{3}}} + 3 e^{\frac{\sqrt{3}\phi}{2}} \right).
\end{equation}

The scalar mass matrix is also given by
\begin{equation}
(m_s^2)^{ij} = \frac{\partial^2 V}{\partial \phi_i \partial \phi_j} = -\frac{1}{4L^2} \left( e^{-\frac{\phi}{\sqrt{3}}} + e^{\frac{\phi}{\sqrt{3}}} \right).
\end{equation}

The negative mass squared indicates that the dilaton $\phi$ is a tachyon in the $3+1$-charge black hole background, and $\phi=0$ is a local maximum and unstable. The actual stable vacua of $\mathcal{N}=8$ supergravity are at other stationary points of the potential, like the $\mathcal{N}=2$ or $\mathcal{N}=1$ vacua, where the scalar masses are positive. So the behavior of this scalar mass matrix is consistent with the stability analysis here. Also, this negative mass square of dilaton would lead to running of the dilaton, so this way the gap, the running of the dilaton and the geometry of the near-horizon region are connected. It also explains the existence of the oscillatory region and the instability of the 3-charge black brane.

\subsection{Composite connections and scalar kinetic terms}

Now here we further study the key mathematical objects needed to construct the covariant derivative for the fermions to find the full Lagrangian. The local $SO(8)$ fundamental gauge fields can be defined as
\begin{gather}
{\mathcal{B}^i}_{\mu \ j}= \frac{2}{3} ({u^{jk}}_{IJ} \partial_\mu {u_{jk}}^{IJ}-v^{ikIJ} \partial_\mu v_{jkIJ}),
\end{gather}
which is the composite $SU(8)$ gauge connections, and it provides the $SU(8)$ part of the covariant derivative.  It ensures that the fermions transform covariantly under local $SU(8)$ transformations, even though the gauge group of the theory is only $SO(8)$. It is also a goldstone boson of the spontaneously broken $E_{7(7)}$ symmetry and encodes the scalar field's kinetic terms and their couplings to fermions.

 The part of $E_7$ which is orthogonal to $SU(8)$ defines another quantity $\mathcal{A}_\mu$,
\begin{gather}
{\mathcal{A}_\mu}^{ijkl}= -2\sqrt{2} (u^{ij}_{IJ}\partial_\mu v^{klIJ}-v^{ijIJ} \partial_\mu u^{kl}_{IJ}),
\end{gather} 
is a non compact part of $E_{7(7)}$ connection, which has the self-duality property
\begin{gather}
{\mathcal{A}_\mu}^{ijkl}= \frac{1}{24} \eta \epsilon^{ijklmnpq} \mathcal{A}_{\mu mnpq}.
\end{gather} 
The parameter $\eta=\pm 1$ is related to the sign of the Chern-Simons term. Note that $A_\mu$ appears in the scalar kinetic terms and in the Pauli couplings between fermions and field strengths, and it encodes the fact that the scalar fields are not free. They actually interact with themselves and with the gauge fields through the non-linear sigma model on the $E_{7(7)}/SU(8)$ coset.

The derivative $D_\mu$ with respect to local $SU(8)$ transformation is
\begin{gather}\label{eq:covdev}
D_\mu \epsilon^i= \partial_\mu \epsilon^i -\frac{1}{2} \omega_{\mu ab}\sigma^{ab} \epsilon^i +\frac{1}{2} \mathcal{B}_{\mu \ j} ^{i} \epsilon^j,
\end{gather}
which defines the covariant derivative acting on the local supersymmetry parameter $\epsilon^i$. Note that the third term here is the composite $SU(8)$ connection, which ensures covariance under local $SU(8)$ transformations. In addition, the supersymmetry parameter $\epsilon^1$ carries an $SU(8)$ index $i$, and so it must transform under both spacetime Lorentz transformation and local $SU(8)$ transformation. This covariant derivative \ref{eq:covdev}, is needed to write down the supersymmetry transformations of the fields.

Also, we have
\begin{gather}
{\omega_\mu}^{ab}=\frac{1}{2} {e_\mu}^c ({\Omega_{ab}}^c-{\Omega_{bc}}^a-{\Omega_{ca}}^b),
\end{gather}
where
\begin{gather}
{\Omega_{ab}}^c= e^\mu_a e^\nu_b(\partial_\mu e^c_\nu-\partial_\nu e^c_\mu -\bar{\psi}^i_{[\mu} \gamma^c \psi_{\nu]i}-\frac{1}{12} \epsilon_{\mu\nu cd} {\bar{\chi}}^{ijk} \gamma^d \chi_{ijk}.
\end{gather}
However, the covariant derivative which acts on $SO(8)$ tensors get another term $g {A_\mu}^{IJ} $. For example
\begin{gather}
D_\mu {u_{ij}}^{IJ}=\partial_\mu {u_{ij}}^{IJ}+\mathcal{B}^k_{\mu [ i} {u_{j] k}}^{IJ}-2g {A_\mu}^{K[I} {u_{ij}}^{J]K}.
\end{gather}
In the above relation, the third term is the fundamental $SO(8)$ gauge field, which acts on the $IJ$ indices, (the $SO(8)$ inices).

The mass term of the Lagrangian is
\begin{gather}
e^{-1} \mathcal{L}_{\text{ferm.mass}}=g {{A_1}^b}_a {\bar{\psi}}_{\mu b} \Gamma ^{ \mu \nu} {\psi^a}_\nu + g {A_{2 \ a}}^m \bar{\chi}_m \Gamma^\mu {\psi^a}_\mu +g {A_{3m}}^n \bar{\chi}_n \chi^m,
\end{gather}
where the first term is the gravitino mass term which gives a mass to the 8 gravitini. This term is specifically crucial for the AdS superalgebra as in AdS spacetime, gravitini can acquire a mass through the super-Higgs mechanism. The second term is the fermion-gravitini mixing term, and it generates mixing between the spin-1/2 and spin-3/2 fields.

The third term also gives a mass to the 56 spin-1/2 fermions, and $A_3$ encodes the fermion mass spectrum. Again, note that these fermion masses are not constant as they depend on the scalar fields (or $r$), which leads to running mass of fermions affecting the fermionic response and the gap.

In each of these steps, one could also find one-loop quantum corrections and therefore find quantum corrections of the gap and fermionic response.

\section{Dirac equation in the background of $4d$ $U(1)^4$ gauged supergravity black brane}
For studying the response of fermionic modes in the background of the assumed geometry, we then study the Dirac equation,
\begin{equation} \label{Diracg}
\Big(i\gamma^\mu \nabla_\mu-m(\phi)+g q_1 \gamma^\mu A_\mu+g q_2 \gamma^\mu a_\mu+ip_1 e^{\frac{\phi}{2\sqrt{3}}} F_{\mu\nu} \gamma^{\mu\nu}+ip_2 e^{-\frac{\sqrt{3}}{2}\phi} f_{\mu\nu} \gamma^{\mu\nu}\Big) \chi=0,
\end{equation}
where the covariant derivative is 
 \begin{gather}
 \nabla_\mu=\partial_\mu-\frac{1}{4} \omega_{\hat{a}\hat{b}\mu} \Gamma^{\hat{a}\hat{b}}.
 \end{gather}
 
 In \ref{Diracg}, the third term couples the fermions to the gauge field $A_\mu$ which is associated with the three equal charge $q$. The fourth term couples the fermions to the gauge field $a_\mu$, which is associated with the single different charge $q'$. The fifth term is the non-minimal coupling where couples fermions directly to the field strength $F_{\mu \nu}$ of the first gauge field, and the exponential factor $e^{\phi/(2\sqrt{3})}$ comes from the supergravity embedding. Similarly, the sixth term is a Pauli coupling to $f_{\mu\nu}$ with a different exponential factor.
 
 These two Pauli terms are important as they break the minimal coupling and allow the fermions to feel the electric fields directly. In holographic Fermi liquid studies, they are crucial for generating the oscillatory region and the gap. So, when finding quantum corrections of the gap, considering the quantum corrections to these term are essential.

We choose the following Gamma matrices,
 
\[                                                          
\Gamma^{\hat{r}}=
  \begin{pmatrix}
    i\sigma_3 & 0 \\
0 & i\sigma_3 \\
  \end{pmatrix},
\]   \[
\Gamma^{\hat{t}}=
  \begin{pmatrix}
    \sigma_1 & 0 \\
0 & \sigma_1 \\
  \end{pmatrix},
\]
 \[
\Gamma^{\hat{i}}=
  \begin{pmatrix}
    i\sigma_2 & 0 \\
0 & -i\sigma_2 \\
  \end{pmatrix},
\]
\\
where the following relationships among them hold,
\begin{gather}
 \{ \Gamma^\mu,\Gamma^\nu\}=2\eta^{\mu\nu},\ \ \ \ \ \ \ \ \ 
 \Gamma^{\mu\nu}=\frac{1}{2}[\Gamma^\mu,\Gamma^{\nu}].
 \end{gather}
The vierbeins, $e_\mu^a$ and the spin connections, $(\omega_{\mu\nu})_a=(e_\mu)_b \Delta_a(e_\nu)^b$ are as follows
\begin{gather}
e_t^a=\frac{1}{\sqrt{g_{tt}}} (\frac{\partial}{\partial t})^a=H(r)^{\frac{1}{4}} f(r)^{-\frac{1}{2}}(\frac{\partial}{\partial t})^a ,\ \ \ \ \ \ 
e_r^a=\frac{1}{\sqrt{g_{rr}}} (\frac{\partial}{\partial r})^a= H(r)^{-\frac{1}{4}} f(r)^{ \frac{1}{2} }  (\frac{\partial}{\partial r})^a , \nonumber\\
e_i^a=\frac{1}{\sqrt{g_{xx}}} (\frac{\partial}{\partial x^i})^a= H(r)^{-\frac{1}{4}}  r^{-1} (\frac{\partial}{\partial x^i})^a,
\end{gather}
\begin{gather}
(\omega_{tr})_t=-(\omega_{rt})_a=-\frac{  \partial_r \sqrt{g_{tt}}   }{\sqrt{g_{rr}}} (dt)_a=(\frac{1}{4} H^{-\frac{3}{2}}\frac{\partial H}{\partial r} f(r)-\frac{1}{2} H^{-\frac{1}{2}}\frac{\partial f}{\partial r})(dt)_a ,\nonumber\\
(\omega_{ir})_a=-(\omega_{ri})_a=\frac{\partial_r \sqrt{g_{xx}}}{\sqrt{g_{rr}}}(dx_i)_a=f^{\frac{1}{2}}(1+\frac{r}{4H})(dx^i)_a .
\end{gather}

Here, one could note that $\frac{1}{4}H^{-3/2}\partial_r Hf-\frac{1}{2}H^{-1/2}\partial_r f$ in $(\omega_{tr})_t$ determines the coupling between the time and radial directions and it encodes the gravitational redshift, and the acceleration of the black hole background. The term $f^{1/2} (1+ \frac{r}{4H}) (dx^i)_a$ in $(\omega_{ir})_a$ denotes the coupling between the spatial and radial directions and encodes the warping of the spatial slices. As the spin connections depend on $r$ through $H(r)$ and $f(r)$, so the Dirac equation would have non-trivial $r$ dependence leading to rich structures in the solutions.

Now for canceling the effects of the spin connections we can write 
\begin{gather}\label{eq:chi}
\chi=\Big(\mathrm{Det}(g) \big | _{r=\mathrm{cte}}\Big)^{-\frac{1}{4}} e^{-i \omega t+i k x} \Psi=H ^{-\frac{1}{8}} r^{-1} f^{-\frac{1}{4}} e^{-i\omega t+ik x}\Psi.
\end{gather}
This can help us to eliminate and simplified the $r$ dependent terms.

We can define the projectors similar to \cite{DeWolfe:2012uv} as, 
\begin{equation}\label{eq:projectors}
\Pi_{\alpha}\equiv \frac{1}{2}\Big(1-(-1)^\alpha i \gamma^{\hat{r}} \gamma^{\hat{t}} \gamma^{\hat{i}}\Big), \ \ \ \ \ \ \ \ P_{\pm} \equiv \frac{1}{2}\Big(1\pm i\gamma^{\hat{r}}\Big).
\end{equation}

Here, $\Pi_{\alpha}$projects onto states with a definite helicity, where the parameter $\alpha=0,1$ distinguishes the two helicity sector. Also, $P_{\pm}$ projects onto states with a definite radial chirality.

Then, using \ref{eq:chi}, \ref{eq:projectors}, and the definition for the gamma matrices, we can split the Dirac equation into two equations for $\Psi_{\alpha+}$, and $\Psi_{\alpha-}$ as 
\begin{eqnarray}
\Big(\partial_r-H^{\frac{1}{4}} f(r)^{-\frac{1}{2}} m(\phi)\Big) \Psi_{\alpha+}&=&[-u(r)+(-1)^\alpha k_i r^{-1}  f(r)^{-\frac{1}{2}}, -v(r)] \Psi_{\alpha -} \nonumber\\  \nonumber\\
\Big(\partial_r+H^{\frac{1}{4}} f(r)^{-\frac{1}{2}} m(\phi)\Big) \Psi_{\alpha-}&=&[u(r)+(-1)^\alpha k_i r^{-1} f(r)^{-\frac{1}{2}} -v(r)], \Psi_{\alpha +} 
\end{eqnarray}
where 
\begin{equation}
u(r)\equiv H^{\frac{1}{2}} f(r) ^{-1} (\omega +g q_1 \Phi_1+g q_2 \Phi_2), \ \ \ \ \ \ \  v(r) \equiv 2 \ H^{\frac{1}{4}} f(r) ^{-\frac{1}{2}} ( p_1 e^{\frac{\phi}{2\sqrt{3}}} \partial_r \Phi_1+ p_2 e^{-\frac{\sqrt{3}}{2} \phi} \partial_r \Phi_2).
\end{equation}

These two functions, $u(r)$ and $v(r)$ encapsulates all the couplings to the background fields. Their new horizon behavior determines the Fermi surfaces and the gap.

Then, the Dirac equation becomes two decoupled second order equations as
\begin{equation}\label{eq:diracsplit}
\Psi^{\prime\prime}_{\alpha\pm}-F_{\pm} \Psi^\prime_{\alpha\pm}+\Big[ \mp \partial_r \big(m \ H^{\frac{1}{4}} f(r)^{-\frac{1}{2}}\big) -m^2 H^{\frac{1}{2}}f(r)^{-1} +\big( u(r)^2-\big(v(r)-(-1)^\alpha k_i r^{-1}  f(r)^{-\frac{1}{2}}\big)^2\big) \pm m \ H^{\frac{1}{4}} f(r)^{-\frac{1}{2}} F_{\pm} \Big] \Psi_{\alpha \pm}=0,\nonumber\\
\end{equation}
where
\begin{equation}
F_{\pm}=\partial_r \mathrm{log} \big[u(r) \mp (-1)^\alpha k_i r^{-1}  f(r)^{-\frac{1}{2}} \pm v(r)\big].
\end{equation}

This equation is the master equation for the fermionic fluctuations. It would be very interesting to find one-loop or higher loop corrections to this equation as well. Solving this equation with appropriate boundary conditions gives the retarded Green's function $G_R(\omega,k)$, which encodes all the fermionic response of the system, and the poles of $G_R$ (at $\omega=0$, $k=k_F$) give the Fermi surfaces.

As the structure of $5d$ and $4d$ are similar, we could deduce that the physics of the gap and Fermi surfaces is universal and robust and not an artifact of the dimensionality and it is an important result. We expect that even considering quantum corrections in various dimensions, keep this universality and robustness.

Similar to \cite{DeWolfe:2012uv}, one can see that the solution of \ref{eq:diracsplit} is invariant under
\begin{equation}
p_i\to -p_i, \ \ \ \ \ \  q_i \to -q_i, \ \ \ \ \ \ \omega \to -\omega, \ \ \ \ \ \ \ \  k \to -k,
\end{equation}
So, the solution of the conjugate fermions have a same solution by changing $(k,\omega)\to (-k,-\omega)$. This is in fact a particle-hole symmetry.

Our geometry in the near boundary limit $r\to \infty$ is $\mathrm{AdS}_4$, where the mass term becomes dominant there. Similar to \cite{DeWolfe:2012uv}, if $ \big | mL\big | \neq 1/2$, the solution is
\begin{equation}
\Psi_{\alpha +} \sim A_{\alpha}(k) r^{mL} + B_{\alpha} (k) r^{-mL-1}, \ \ \ \ \ \ \  \Psi_{\alpha-}\sim C_{\alpha}(k)r^{mL-1}+D_{\alpha}(k) r^{-mL}.
\end{equation}
Now, we can see that we have the same relations between the coefficients of the solutions as
\begin{equation}
 C_\alpha= \frac{L^2 (\omega +(-1)^\alpha k)}{2mL-1} A_\alpha, \ \ \ \ \ \ B_\alpha=\frac{L^2(\omega-(-1)^\alpha k)}{2mL+1} D_\alpha,
 \end{equation}
where $mL\equiv 2(m_1+m_2)$. If $m>0$, $A$ would be the source term and $D$ is the response, and it would be vise versa for the case of $m<0$, and for a dual fermion with an opposite chirality \cite{DeWolfe:2013fha}.

For $mL=\frac{1}{2}$, we have
\begin{equation}
 C_\alpha= L^2 (\omega +(-1)^\alpha k) A_\alpha, \ \ \ \ \ \ \ \ \ \ \  B_\alpha=\frac{L^2(\omega-(-1)^\alpha k)}{2mL+1} D_\alpha,
 \end{equation}
and then the retarded Green's function for the dual fermionic operator for the fluctuations at the horizon would be the ratio of the response to the source, as in the following relation
\begin{equation}
(G_R)_{\alpha\beta}=\frac{D_\alpha}{A_{\beta}}.
\end{equation}
Similar to \cite{DeWolfe:2012uv}, our equation \ref{eq:diracsplit}, is being decoupled for the $\Psi_{\alpha\pm}$, so our Green's function is also diagonal. Therefore, we get the relation $G_{22}(\omega, k)=G_{11} (\omega, -k)$. Next, one can find the Fermi surfaces where $k=k_F$. This can be done by finding the poles of the Green's function at Fermi energy which are at zero frequency \cite{Faulkner:2009wj, Belliard:2011qq}, as in the relation
\begin{gather}
A_\beta(\omega=0, k=k_F)\equiv 0.
\end{gather}
The collection of all such poles for different $k_F$ defines the Fermi surface topology. Also, as these conditions for $4d$ and $5d$ cases are similar, we could deduce that this is universal and applies to both $4d$ and $5d$ cases. One could check this for other dimensions and also by considering various loop quantum corrections, and analyze the Fermi surface topology in these cases.

\subsection{Near-horizon analysis: 3+1-charge case}
In this section we study the near horizon limit of the Dirac equation first for the general ``regular" 3+1-charge case and then we look at the near horizon solution of the extremal 3-charge and then 1-charge black brane.

By defining the parameters, 
\begin{gather}
\beta _1=\frac{1}{L\left(q+r_H\right){}^2}\sqrt{\frac{q \big(\left(q+r_H\right){}^4+L^2 q \left(q-2 r_H\right)\big)}{q-2 r_H}}, \ \ \ \ \ \ 
\beta _2=\frac{\left(q-2r_H\right)\sqrt{3 \left(\left(q+r_H\right){}^4+3L^2r_H{}^2\right)}}{L r_H\left(q+r_H\right){}^2}, \nonumber\\
{k_0}^2=\frac{\left(q+r_H\right){}^2 \sqrt{r_H}}{\sqrt{q-2r_H}}, \ \ \ \ \ \ \ \ \  (L_2)^2=\frac{ L^2 \sqrt{r_H (q-2r_H)}}{3 (q-r_H)},
\end{gather}
the leading terms for each component of the metric and the gauge fields of the 3+1-charge ``extremal" supergravity black brane can be detected. Here one can eliminate $q^\prime$ by the extremity condition \ref{eq:extg}, which would lead to
\begin{gather}
g_{tt} =-\frac{1}{ (L_2)^2} (r-r_H)^2+O\left[r-r_H\right]{}^3,\ \ \ 
g_{rr} = (L_2)^2 (r-r_H)^{-2}+O\left[r-r_H\right]{}^{-1}\ \ \  g_{ii} = {k_0}^2+O\left[r-r_H\right]{}^1.
\end{gather}
One can see that, in the general 3+1-charge case, an $\mathrm{AdS}_2$ factor is appeared in the near horizon geometry. This factor is responsible for the emergent conformal symmetry in the IR which leads to non-Fermi liquid behavior and the gap. Also as in $AdS_2/CFT_1$, there is no dynamics, and the IR dynamics are frozen or "gapped", it would be consistent with our result of the presence of a gap in the dual field theory.

The leading terms of the gauge fields then are
\begin{gather}
\Phi_1 =\beta _1 \left(r-r_H\right)+O\left[r-r_H\right]{}^2,\ \ \ \ \ \ 
\Phi_2= \beta _2 \left(r-r_H\right)+O\left[r-r_H\right]{}^2,
\end{gather}
which are linear in $r$ and similar to the 2+1-charge case of $5d$ \cite{DeWolfe:2012uv, DeWolfe:2013fha} have a single zero. The parameters $\beta_1$ and $\beta_2$ and also $p_1$ and $p_2$ are dimensionless.

Also, the leading term of the scalar field $\phi=\sqrt{3}\  \mathrm{log} X$, near the horizon would be a constant and independent of $r$, as
\begin{gather}
\phi = \phi_0=\frac{\sqrt{3}}{4} \log \left(\frac{r_H}{q-2 r_H}\right)+O[r-r_H]^2.
\end{gather}
This constant electric field is responsible for the Pauli couplings in the Dirac equation, and the source of the oscillatory region and the gap.

Now, we study the near horizon limit of the Dirac equation. For doing so, first, in the near horizon limit $(r\to r_H)$, we assume $ \omega=0$, and then we can study the small $\omega$ limits, while taking $\frac{\omega}{r-r_H}$ which is constant, and then we can comment on how these two limits are non-commutative.

So, first, for $ \omega=0$ in the near horizon limit, one finds
\begin{gather}
u = \frac{\left(L_2\right){}^2}{2 L }\left(q \beta _1+\frac{3r_H{}^2}{q-2 r_H} \beta _2\right)\frac{1}{r-r_H}+O\left[r-r_H\right]{}^0 ,\nonumber\\  \nonumber\\ 
 v =2 L_2\left(\text{  }p_1 e^{\frac{ \phi }{2\sqrt{3}}} \beta _1+ p_2 e^{-\frac{\sqrt{3} \phi }{2}} \beta _2\right)\frac{1}{\text{   }r-r_H}+O\left[r-r_H\right]{}^0.
 \end{gather}
Then, the near horizon limit of $F_{\pm}\to F_{NH}$ would be
\begin{gather}
F_{ NH}=-\frac{1}{r-r_H}.
\end{gather}
This form of near-horizon limits of the function $F$ makes the form of Dirac equation in this limit much simpler. Then, the Dirac equation similar to \cite{DeWolfe:2012uv} would simplify to
\begin{gather}
\partial_r ^2 \Psi _{\alpha \pm} +\frac{1}{r-r_H}\Psi _{\alpha \pm}^\prime -\frac{\nu_k ^2}{(r-r_H)^2} \Psi_{\alpha\pm}=0,
\end{gather}
where $\nu_k$ is
\begin{equation} \label{eq:nu}
\begin{split}
\nu _k=\sqrt{\left( m^2(\phi_0 ) +\left(\frac{\tilde{k}}{k_0}\right){}^2\right)\left(L_2\right)^2-\frac{\left(L_2\right){}^4 }{4 L^2} \left( q_1 \beta _1+\frac{3 r_H{}^2 }{ q_1-2r_H} \beta _2\right)^2},
\end{split}
\end{equation}
with 
\begin{gather}
\tilde{k} \equiv k_i-(-1)^{\alpha }2 k_0\left( p_1 e^{\frac{ \phi_0 }{2\sqrt{3}}}\beta _1+\text{  }p_2 e^{-\frac{\sqrt{3}}{2} \phi_0 }\beta _2\right).
\end{gather}

This is the shifted momentum, where this shift is due to the Pauli terms in the Dirac equation. Note that for two helicity sectors of $\alpha=1$ and $\alpha=2$, the shift in momentum is different. So the gap and instability depends on the helicity of the fermion.

Note that as we tried to choose the most similar way of defining the parameters for the sake of comparison, this equation looks similar to the five-dimensional case of \cite{DeWolfe:2012uv}. However, the definitions of the quantities are different from those in \cite{DeWolfe:2012uv}, and there are also some differences between the above equation and the corresponding one in \cite{DeWolfe:2012uv}. For example, $\frac{L_2}{\tau_0}$ is replaced with $L_2^4$, and the definition of $L_2$ here is different from the one in \cite{DeWolfe:2012uv}. However, since the equations show similar behavior in four and five dimensions, the physics that we extract from these formulas would have similarities to \cite{DeWolfe:2012uv}. We can mention a few of those results here again.

First, one can check that the effect of the Pauli terms is just to shift the origin of the 3-momentum $k_i$. However, for $\alpha_1$ and $\alpha_2$, the shifts due to the different signs are in opposite directions. Similarly, the solution to the Dirac equation is of the form
\begin{equation}
\Psi \sim (r-r_H)^{\pm \nu_k},
\end{equation}
and this solution has a hidden near-horizon region approaching $\mathrm{AdS}_2 \times \mathbb{R}^2$. We can see this by defining
\begin{equation}
r - r_H = \lambda (L_2)^2 \frac{1}{\zeta}, \qquad t = \frac{1}{\lambda} \tau.
\end{equation}
Now, by taking the limit $\lambda \to 0$ and keeping $\zeta$ and $\tau$ fixed, we find the near-horizon metric as
\begin{equation}
ds^2 = \frac{(L_2)^2}{\zeta^2} (-d\tau^2 + d\zeta^2) + k_0^2 \, d\vec{x}^2.
\end{equation}
The radius of this $\mathrm{AdS}_2 \times \mathbb{R}^2$ metric is
\begin{equation}
L_{\mathrm{AdS}_2} = L_2 = \frac{L \left(r_H (q - 2r_H)\right)^{1/4}}{\sqrt{3(q - r_H)}}.
\end{equation}

The near-horizon gauge fields are
\begin{equation}
A_\mu dx^\mu = \frac{\beta_1 (L_2)^2}{\zeta} d\tau, \qquad
A_\mu' dx^\mu = \frac{\beta_2 (L_2)^2}{\zeta} d\tau.
\end{equation}
Matching the boundary conditions in and out of the $\mathrm{AdS}_2$ geometry requires one to choose the positive sign of $\nu$, i.e., $\Psi \sim (r - r_H)^{+\nu_k}$.

The oscillatory region is where $\nu_{k_{\text{osc}}}$ is imaginary \cite{DeWolfe:2012uv}. This happens when the effective electric coupling
\begin{equation}
(qe)_{\text{eff}} \equiv \frac{(L_2)^2}{2L} \left( q_1 \beta_1 + \frac{3 r_H^2}{q_1 - 2r_H} \beta_2 \right)
\end{equation}
is stronger than the effect of the mass and the shifted momentum $\tilde{k}$. This is intuitively correct, since a higher mass of the scalar field near the horizon, which is associated with larger chemical potentials $\mu_1$ and $\mu_2$, can make the near-horizon region more stable. Additionally, a higher momentum $k_i$ of infalling waves is associated with greater stability of the states. On the other hand, larger electric couplings near the horizon can make this region more unstable, which would lead to particle creation in the near-horizon $\mathrm{AdS}_2$ region.

As mentioned in \cite{DeWolfe:2012uv}, the fact that the mass term here is a function of the scalar field, and consequently a function of the chemical potentials $\mu_i$, is specifically a feature of solutions arising from gauged supergravity theories in the top-down approach \cite{Chow:2013gba}.

One can now find where the oscillatory region appears by using \eqref{eq:nu} and solving
\begin{equation}
\nu_{k_{\text{osc}}} = 0.
\end{equation}
Similar to \cite{DeWolfe:2012uv}, for the four-dimensional case this happens where
\begin{equation}
\tilde{k}_{\text{osc}}^2 = \left(\frac{k_0}{L_2}\right)^2 \left( (qe)_{\text{eff}}^2 - m^2(\phi) L_2^2 \right).
\end{equation}

This oscillatory momentum can exist only when the effective near-horizon electric field coupling is greater than the effective mass term. This is also corresponds to pair creation and Schwinger effects, and also non-Fermi liquid behavior and the gap in the dual field theory. As the relation is the same for $4d$ and $5d$, this is a genuine holographic phenomenon.

The quantum correction of this relation $\delta \tilde{k}^2_{\text{osc}}$, get contributions from higher curvature corrections, higher order gauge field corrections, and fermion 1-loop effective action, where each contribution can be found in future works.

We now investigate the existence of the oscillatory region in the near-horizon geometry of the two cases of 3-charge ($q' \to 0$) and 1-charge ($q \to 0$) black brane solutions.

The near-horizon oscillatory momentum $k_{\text{osc}}^2$ for the 3-charge black brane ($q' \to 0$) is
\begin{equation}
k_{\text{osc}}^2 = \frac{\sqrt{3q'} \, q^{3/2}}{36 L^2} \left( L^2 + q^2 \right) 
-\frac{q^{5/4}}{2L^2} \Bigg( m_1 m_2 \left(\frac{q'}{3}\right)^{1/4} \left( \frac{5}{2} \sqrt{\frac{q'}{3}} + \sqrt{q} \right) 
+ \frac{m_2^2 q^{3/4}}{2} 
+ \frac{q^{1/4} \sqrt{3q'}}{6} \left( m_1^2 + 2m_2^2 \right) \Bigg) 
+ \mathcal{O}(q')^1.
\end{equation}
One can see that for the oscillatory region to exist, $m_1$ and specifically $m_2$ should be smaller than $q$. The effect of $m_2$ is more important than that of $m_1$ due to the third and fourth terms. Thus, in this case, $q$ and $m_2$ are competing with each other for the positivity of the right-hand side and therefore for the existence of $k_{\text{osc}}$.

The oscillatory region cannot exist for the 1-charge black hole, as the leading term in the expansion of $k_{\text{osc}}$ for the case of $q \to 0$ is negative:
\begin{equation}
k_{\text{osc}}^2 = -\frac{3 m_1^2 q q'}{8 L^2} - \frac{3 \sqrt{3q'} \, q^{3/2} m_1 m_2}{4\sqrt{2} L^2} + \frac{q^2}{432 L^2} \left( 169 L^2 + 81 m_1^2 - 243 m_2^2 \right) + \mathcal{O}(q)^{5/2}.
\end{equation}
As one can see, the right-hand side is always negative, and therefore, for the four-dimensional 1-charge black brane, the near-horizon geometry is always stable.

Now we look for the case where $\omega$ is non-zero to study its effects on the Dirac equation and the behavior of the dispersion relation. To do so, we turn to the notation of \cite{DeWolfe:2013fha}. By defining $U_{\pm}$, one can write the four-dimensional near-horizon Dirac equation \ref{Diracg} as
\begin{equation}
U^{\prime\prime} + \left( \frac{1}{r - r_H} + \cdots \right) U^{\prime}
+ \left( \frac{L^4 (q - 2r_H) r_H \omega^2}{9 (q - r_H)^2 (r - r_H)^4}
+ \frac{n \omega}{(r - r_H)^3}
- \frac{\nu^2}{(r - r_H)^2} + \cdots \right) U = 0,
\end{equation}
where
\begin{equation}
n = \frac{i L^2 \sqrt{r_H (q - 2r_H)}}{3 (q - r_H)}
+ \frac{L^3 r_H \left( q \beta_1 (q - 2r_H) + 3 \beta_2 r_H^2 \right)}{9 (q - r_H)^2},
\end{equation}
and $\nu^2$ is a complicated quantity that depends on $\omega^2$, $\omega$, and a term that depends only on $r_H$, $L$, and $p$. For $\omega = 0$, only the $\nu^2$ term remains, and the $\frac{1}{r - r_H}$ term becomes dominant.
The slope of the dispersion relation near the Fermi surface gives the Fermi velocity, which determines the transport properties of the system.

The hierarchy of singularities is important as the most singular term $(1/\zeta^4)$ dominates near the horizon, but the $(1/\zeta^3)$ and $(1/\zeta^2)$ terms become important at intermediate distances.

One could also note that the first term of $n$ is purely imaginary, and comes from the $i$ in the Dirac equation and is related to the infalling boundary condition at the horizon. So it introduces dissipation into the system which corresponds to the finite lifetime of quasiparticles in the dual field theory. The second term of $n$ is real and comes from the gauge field couplings. It encodes the electric field effects near the horizon. The imaginary part of $n$ leads to a complex dispersion relation $\omega(k)$, which means the quasiparticles have a finite lifetime. This structure again is similar to $5d$ case and therefore this finite-frequency analysis is universal and does not depend on the spacetime dimension, as $n$ always encodes electric filed effects and the dissipation.

\section{Uplifting the metric to five dimensions}
First, by taking the near-horizon ($r \to 0$) limit of the $4d$ extremal 3-charge black brane metric, we obtain
\begin{equation}\label{eq:nearh}
ds^2 = \frac{q^{5/2}}{r^{1/2} L^2} dt^2 - \frac{r^{1/2} L^2}{q^{5/2}} dr^2 + q^{1/2} r^{3/2} d\vec{x}^2.
\end{equation}
The scalar field behaves near the horizon as
\begin{equation}
e^{\frac{\phi}{2\sqrt{3}}} = \left( \frac{q}{r} \right)^{1/4}.
\end{equation}
The Kaluza–Klein reduction ansatz for a $5d$ metric that, after reduction, becomes \eqref{eq:nearh} is
\begin{equation}
d\hat{s}^2 = e^{2\alpha \phi} ds^2 + e^{2\beta \phi} (dz + \mathcal{A})^2
= e^{-5\phi/\sqrt{3}} ds^2 + e^{-4\phi/\sqrt{3}} (L \, d\varphi_3 + \mathcal{A})^2.
\end{equation}
Then the $5d$ metric becomes
\begin{equation}
d\hat{s}^2 = \frac{r^2}{L^2} dt^2 - \frac{L^2}{r^2} dr^2 + \frac{r^2}{q^2} L^2 d\varphi_3^2 + \frac{r^4}{q^2} d\vec{x}_{2,k}^2.
\end{equation}

This $5d$ metric is smooth and is clearly recognized as $\text{AdS}_3 \times$ (a compactified $2d$ space), as it has an $\mathrm{AdS}_3$ or a null-warped $\mathrm{AdS}_3$ sector. So the gap we discussed before is now mapped to this decoupled $\text{AdS}_3$ sector in the $5d$ spacetime.

It would be interesting to find the central charges of the dual $\mathrm{CFT}_2$ and study its Virasoro algebra. As it has a null Killing vector, the symmetry algebra of the deformed $\text{AdS}_3$ would reduce to a single copy of the Virasoro algebra plus a $U(1)$ current algebra.

\subsection{Near-horizon limit in $11d$}

The geometry of the regular 3+1-charge black brane in $11d$ is
\begin{align}
ds_{11}^2 &= \Delta^{2/3} \left[ -\frac{f}{H} dt^2 + \frac{dr^2}{f} + r^2 (dx^2 + dy^2) \right] + ds_7^2, \label{eq:11d}\\
ds_7^2 &= \Delta^{-1/3} \Bigg[ L^2 \mathcal{H} \left( \sum_{i=1}^3 d\mu_i^2 + \mu_i^2 \left( d\phi_i + A_i \frac{dt}{L} \right)^2 \right) \nonumber\\
&\qquad + L^2 \mathcal{H}' \left( d\mu_4^2 + \mu_4^2 \left( d\phi_4 + A' \frac{dt}{L} \right)^2 \right) \Bigg],
\end{align}
where
\begin{equation}
H_1 = H_2 = H_3 = \mathcal{H} = 1 + \frac{q_1}{r}, \qquad
H_4 = \mathcal{H}' = 1 + \frac{q_2}{r}, \qquad
H = \mathcal{H}^3 \mathcal{H}'.
\end{equation}

If, using equation \eqref{eq:11d}, one considers $\mu_4 = \cos\theta$, the general form of $\Delta$ in \cite{Fareghbal:2008dy} can be written as
\begin{equation}
\Delta = \mathcal{H}' \mathcal{H}^3 \left( \frac{1 - \mu_4^2}{\mathcal{H}} + \frac{\mu_4^2}{\mathcal{H}'} \right)
= \mathcal{H}^2 \left[ \mathcal{H}' + \mu_4^2 (\mathcal{H} - \mathcal{H}') \right]
= \mathcal{H}^2 \left[ \mathcal{H}' + \mu_4^2 \left( \frac{q_1 - q_2}{r} \right) \right].
\end{equation}

The three-form field strength $C_3$, satisfying $F_4 = dC_3$, is \cite{Fareghbal:2008dy}
\begin{align}
C^{(3)} &= -\frac{r^3}{2} \Delta \, dt \wedge d^2 \sigma_2 \nonumber\\
&\quad - \frac{L^2}{2} \sum_{i=1}^3 Q_i \mu_i^2 \left( d\phi_i - \frac{q_i}{Q_i} \frac{dt}{L} \right) \wedge d^2 \Omega_2 \nonumber\\
&\quad - \frac{L^2}{2} \left( Q_4 \mu_4^2 \left( d\phi_4 - \frac{q_4}{Q_4} \frac{dt}{L} \right) \wedge d^2 \Omega_2 \right),
\end{align}
where $d^2 \Omega_2$ is the volume form on a unit-radius two-sphere.

Now we consider two cases for taking the near-horizon limits. In both of these cases, $q_1$ is a finite non-zero value.

In the first scenario, we first consider $q_2 \to 0$, and then we take the near-horizon limit $r \to r_H$. As shown in equation \eqref{eq:ext1}, for the non-BPS black hole, the horizon is at $r_H = 0$. So in this case, we will obtain the near-horizon geometry of the 3-charge black hole in $11d$ at the end. We call this limit $L_1$.

In the other case, while keeping $q_2$ finite (which corresponds to a finite horizon $r_H$), we take the near-horizon limit $r \to r_H$, which leads to the near-horizon geometry of the 3+1-charge geometry in $11d$. In the next step, we can consider $q_2 \to 0$ while keeping the solution in the near-horizon geometry. We call this limit $L_2$.

In brief, the orders of the limits are as follows:
\begin{align}
L_1 &: \quad q_2 \to 0, \quad r \to r_H, \quad r_H \to 0, \\
L_2 &: \quad q_2 = \text{finite}, \quad r \to r_H, \quad r_H = \text{finite}, \quad q_2 \to 0.
\end{align}

Now we study each scenario in more detail.

\subsection{Near horizon limit in $11d$}

The geometry of regular 3+1-charge black brane in $11d$ is
\begin{gather}\label{eq:11d}
ds_{11}^2=\Delta^{\frac{2}{3}}\Big[-\frac{f}{H} dt^2+\frac{dr^2}{f}+r^2 (dx^2+dy^2)\Big]+{ds_7^2}\nonumber\\
ds_7^2=\Delta^{-\frac{1}{3}}\bigg[L^2 \mathcal{H} \Big(\sum\limits_{i=1}^3 d\mu_i^2+\mu_i^2\big(d\phi_i+A_i\frac{dt}{L}\big)^2\Big)+L^2 \mathcal{H}^\prime\Big(d\mu_4^2+\mu_4^2\big(d\phi_4+A^\prime\frac{dt}{L}\big)^2\Big)\bigg],
\end{gather}
where
\begin{gather}
H_1= H_2=H_3=\mathcal{H}=1+\frac{q_1}{r}, \ \ \ \ \ \ \ \
H_4=\mathcal{H^\prime}=1+\frac{q_2}{r},\ \ \ \ \ \ \ \ \ 
H=\mathcal{H}^3 \mathcal{H^\prime}.
\end{gather}

If by using equation \ref{eq:11d}, one considers $\mu_4=\text{cos}\ \theta$, the general form of $\Delta$ in \cite{Fareghbal:2008dy} could be written as
\begin{gather}
\Delta= \mathcal{H^\prime}\mathcal{H}^3\Big(\frac{1-\mu_4^2}{\mathcal{H}}+\frac{\mu_4^2}{\mathcal{H}^\prime}\Big)= \mathcal{H}^2\big[\mathcal{H}^\prime+\mu_4^2\big(\mathcal{H}-\mathcal{H^\prime}\big)\big]=
\mathcal{H}^2\big[\mathcal{H}^\prime+\mu_4^2\big(\frac{q_1-q_2}{r}\big)\big].
\end{gather}
The three form-field strength $C_3$ satisfying $F_4=dC_3$ is\cite{Fareghbal:2008dy}
\begin{gather}
C^{(3)}=-\frac{r^3}{2} \Delta dt \wedge d^2 \sigma_2 -\frac{L^2}{2} \sum\limits_{i=1}^3 Q_i \mu_i^2 \big(d\phi_i-\frac{q_i}{Q_i}\frac{dt}{L}\big)\wedge d^2 \Omega_2-\frac{L^2}{2} \Big(Q_4\mu_4^2\big(d\phi_4-\frac{q_4}{Q_4}\frac{dt}{L}\big)\wedge d^2\Omega_2\Big),
\end{gather}
where, $d^2\Omega_2$ is the volume form on a unit radius two sphere.

 Now we consider two cases for taking the near horizon limits. In both of these cases, $q_1$ is a finite non-zero value. 
In the first scenario, we first consider $q_2 \to 0$, and then we take the near horizon limit $r\to r_H$. As it has been shown in the equation \ref{eq:ext1}, for the non-BPS black hole, the horizon is at $r_H=0$. So in this case, we will get the near horizon geometry of 3-charge black hole in $11d$ at the end. We call this limit $L_1$.

Then for the other case, while keeping $q_2$ finite, which corresponds to a finite horizon $r_H$, we take the near horizon limit $r \to r_H$, which will lead to the near horizon geometry of 3+1-charge geometry in $11d$. In the next step, we can consider  $q_2 \to 0$ while keeping the solution in the near horizon geometry. We call this limit $L_2$.

 So in brief the orders of the limits are as follows \\ \\
$L_1$: \ \ $q_2 \to 0$, \ \ \ \ \ \ \           $r\to r_H$,  \ \ \ \ $r_H\to 0$\\
$L_2$: \ \ $q_2=\text{finite}$, \ \ \ $r\to r_H$, \ \ \ \ $r_H=\text{finite}$, \ \ \ \ $q_2 \to 0$ \\

Now we study each scenario in more details.

\subsubsection{$L_1$ Limit}

For the $L_1$ case, we have
\begin{align}
f &= -\frac{\mu}{r} + \frac{r^2}{L^2} \left( 1 + \frac{q_1}{r} \right)^3 \left( 1 + \frac{q_2}{r} \right) \nonumber\\
&\simeq \frac{3q_1^2}{L^2} + \frac{1}{r} \left( -\mu + \mu_{cr} + \frac{3q_1^2 q_2}{L^2} \right) + \frac{q^3 q_2}{r^2 L^2} : \text{Finite}, \qquad \mu_{cr} = \frac{q_1^3}{L^2}.
\end{align}
In order to keep $f$ finite, we need to consider the following relations
\begin{equation}
r \sim r_H \sim \epsilon, \qquad
q_2 \sim \epsilon^2, \qquad
(\mu - \mu_{cr}) \sim \epsilon.
\end{equation}
Therefore, we get
\begin{align}
f &\simeq \frac{3q_1^2}{L^2} - \frac{\mu - \mu_{cr}}{r} + \frac{q_1^3 q_2}{r^2 L^2}, \\
\mathcal{H} &= 1 + \frac{q_1}{r} \simeq \frac{q_1}{r}, \qquad
\mathcal{H}' = 1 + \frac{q_2}{r} \simeq 1 + \mathcal{O}(\epsilon), \qquad
\Delta \simeq \mu_4^2 \frac{q_1^3}{r^3}, \\
A &= \frac{Q_1}{q_1} \left( \frac{1}{\mathcal{H}} - 1 \right) \simeq -\frac{Q_1}{q_1} \simeq -\sqrt{1 + \frac{q_1^2}{L^2}}, \\
a &= \frac{Q_2}{q_2} \left( \frac{1}{\mathcal{H}'} - 1 \right) \simeq -\frac{Q_2}{r} \simeq -\sqrt{\frac{q_1^3}{L^2}}.
\end{align}

The above limits in the parameter space should be accompanied by the following near-horizon limit over the coordinates
\begin{equation}
(r - r_H) \sim \epsilon, \qquad
t \sim \epsilon^{-1/2}, \qquad
\phi_4 \sim \epsilon^{-1/2}.
\end{equation}

Using these, we can find the limits of the components of the metric
\begin{align}
r^2 \Delta^{2/3} d\Omega_2^2 &= q_1^2 \mu_4^{4/3}, \\
-\Delta^{2/3} \frac{f}{H} dt^2 &= -\mu_4^{4/3} \frac{r}{q_1} f \, dt^2, \\
\frac{\Delta^{2/3}}{f} dr^2 &= \mu_4^{4/3} \frac{q_1^2}{f} \frac{dr^2}{r^2} : \text{Finite}.
\end{align}
Note that $\frac{dr}{r}$ is finite, because if we consider $r = r_H + \rho$, then $\frac{d\rho}{(r_H + \rho)}$, and therefore $g_{rr} = \mu_4^{4/3} \frac{q_1^2}{f} \left( \frac{d\rho}{\rho + r_H} \right)^2$, would be finite.

Now, considering
\begin{equation}
d\phi_i + A_i \frac{dt}{L} = d\phi_i - \sqrt{1 + \frac{q_1^2}{L^2}} \frac{dt}{L} = d\varphi_i,
\end{equation}
the components of the $ds_7$ part are
\begin{align}
\mu_i, \phi_i &: \quad \mu_4^{-2/3} L^2 \sum_{i=1}^3 \left( d\mu_i^2 + \mu_i^2 d\varphi_i^2 \right), \\
\mu_4, \phi_4 &: \quad \mu_4^{-2/3} \frac{r}{q_1} L^2 \left( d\mu_4^2 + \mu_4^2 \left( d\phi_4 - \sqrt{\frac{q_1^3}{L^2}} \frac{dt}{L} \right)^2 \right).
\end{align}
So finally, the near-horizon geometry in this case is
\begin{align}
ds^2 &= \mu_4^{4/3} \Bigg[ -\frac{r}{q_1} f \, dt^2 + \frac{q_1^2}{r^2} \frac{dr^2}{f} + \frac{r}{q_1} L^2 \left( d\phi_4 - \sqrt{\frac{q_1^3}{L^2}} \frac{dt}{L} \right)^2 + q_1^2 (dx^2 + dy^2) \Bigg] \nonumber\\
&\quad + \mu_4^{-2/3} L^2 \sum_{i=1}^3 \left( d\mu_i^2 + \mu_i^2 d\varphi_i^2 \right).
\end{align}
As can be seen, in this order of limits, the near-horizon geometry is $\mathrm{AdS}_3 \times T^2 \times T^6$ or $\mathrm{BTZ} \times T^2 \times T^6$.

The BTZ sector is rotating with angular momentum $\frac{\sqrt{q_1^3}}{L^2}$.
For the case $q_1 = -r_H$ and therefore $\mu = 0$ (the BPS point), the three-dimensional part describes a global $\mathrm{AdS}_3$ space \cite{Fareghbal:2008dy}.

The scalar fields in this limit are
\begin{equation}
X \simeq \left( \frac{r}{q_1} \right)^{1/4} \simeq 0, \qquad
X' \simeq \left( \frac{q_1}{r} \right)^{3/4} \simeq q_1^{3/4} \epsilon^{-3/4}.
\end{equation}
Therefore, by considering the limit $q_2 \to 0$ and then taking the near-horizon limit, only one scalar field remains present in the near-horizon geometry, which is $\mathrm{AdS}_3 \times T^2 \times T^6$.

\subsubsection{$L_2$ Limit}

Now, for the other case $L_2$, both $q_2$ and $r_H$ are finite. After taking the near-horizon limit $r \to r_H$, the components of the metric are
\begin{align}
g_{tt} &= -\frac{3(q_1 - r_H)}{L^2 r_H} \left( \frac{r_H + \mu_4^2 (q_1 - 3r_H)}{q_1 - 2r_H} \right)^{2/3} (r - r_H)^2 + \mathcal{O}[(r - r_H)^3], \\
g_{rr} &= \frac{L^2 (q_1 - 2r_H)^{1/3} \left( r_H + \mu_4^2 (q_1 - 3r_H) \right)^{2/3}}{3 (q_1 - r_H) (r - r_H)^2} + \mathcal{O}[(r - r_H)^{-1}], \\
g_{ii} &= \left( \frac{r_H + \mu_4^2 (q_1 - 3r_H)}{q_1 - 2r_H} \right)^{2/3} (q_1 + r_H)^2 + \mathcal{O}[(r - r_H)^1].
\end{align}

For taking the limit of the $ds_7$ part, one needs the limit of the gauge fields
\begin{align}
\Phi_1 &= -\sqrt{ \frac{q_1 \left( L^2 q_1 (q_1 - 2r_H) + (q_1 + r_H)^4 \right)}{L^2 (q_1 - 2r_H) (q_1 + r_H)^2} } + \mathcal{O}[(r - r_H)^1], \\
\Phi_2 &= -\frac{\sqrt{3 \left( (q_1 + r_H)^4 + 3L^2 r_H^2 \right)}}{L (q_1 + r_H)} + \mathcal{O}[(r - r_H)^1],
\end{align}
which are constant in the leading term and are independent of $r$. Therefore, we can simply rename the angular component, $d\varphi_i = d\phi_i + A_i \frac{dt}{L}$, and so the limits of the $ds_7$ part are
\begin{align}
\Delta^{-1/3} L^2 \mathcal{H} \sum_{i=1}^3 d\mu_i^2 &= \frac{L^2 (q_1 - 2r_H)^{1/3}}{ \left( r_H + \mu_4^2 (q_1 - 3r_H) \right)^{1/3} } + \mathcal{O}[(r - r_H)^1], \\
\Delta^{-1/3} L^2 \mathcal{H}' \sum_{i=1}^3 d\mu_i^2 &= \frac{L^2 r_H}{(q_1 - 2r_H)^{2/3} \left( r_H + \mu_4^2 (q_1 - 3r_H) \right)^{1/3}} + \mathcal{O}[(r - r_H)^1].
\end{align}
So all the components of $ds_7$ and also $g_{ii}$ are independent of $r$, and therefore the near-horizon geometry is $\mathrm{AdS}_2 \times T^2 \times T^7$. Then, considering $q_2 \to 0$ imposes $r_H \to 0$.

\section{Conclusion}

In this work, we considered the $U(1)^4$ charged black brane solution of four-dimensional gauged supergravity and examined the fermionic response in this geometry. This geometry is holographically dual to $3d$ $\mathcal{N}=2$ SCFT ABJM models. Similar to \cite{DeWolfe:2013fha}, we showed that a gap exists in the states of the conformal field theory, which corresponds to the different limiting behaviors of the two unequal chemical potentials of the four-charge geometry. We analyzed the behavior of this gap and also the stability of the near-horizon geometry by changing the parameters of the theory in various orders. We categorized the $56$ fermion modes of the geometry and found the coefficients of the Dirac equations for each mode. We then uplifted the geometry to five-dimensional and then to eleven-dimensional geometries and, similar to \cite{Fareghbal:2008dy}, we showed that in both cases, a piece of $\mathrm{BTZ} \times \mathrm{S}^2$ or $\mathrm{AdS}_3 \times \mathbb{R}^2$ emerges, and as a result, a decoupling sector exists in the field theory.

We specifically showed that in $4d$ and $11d$, the order of limits also matters. If we first take $q' \to 0$ and then $r \to r_H$, corresponding to the 3-charge black brane, we obtain an $\mathrm{AdS}_3$ sector in the near-horizon geometry, while if we first take $r \to r_H$ with finite $q'$ and then $q' \to 0$, corresponding to the 3+1-charge black brane, we obtain an $\mathrm{AdS}_2$ sector.

On the gravity side, we also further considered three different limits to examine the black brane in the bulk. These correspond to three different regimes of energy in the dual CFT. Defining $\frac{r - r_H}{r_H} = \delta$ and $q' = \tilde{q}' \epsilon$, the three regimes are:
\begin{align}
\frac{\delta}{\epsilon} > 1 \quad &\text{or} \quad \frac{\tilde{q}'}{q'} \left( \frac{r}{r_H} - 1 \right) > 1 \quad \to \quad \text{system of 3-charge black brane}, \\
\frac{\delta}{\epsilon} < 1 \quad &\text{or} \quad \frac{\tilde{q}'}{q'} \left( \frac{r}{r_H} - 1 \right) < 1 \quad \to \quad \text{3+1-charge black brane}, \\
\frac{\delta}{\epsilon} = 1 \quad &\text{or} \quad \frac{\tilde{q}'}{q'} \left( \frac{r}{r_H} - 1 \right) = 1 \quad \to \quad \text{excitations of $\mathrm{AdS}_3$, BTZ}.
\end{align}

The physical interpretation of this gap and the corresponding energy limits in the bulk can be understood, for instance, by considering a strong electric field turned on in a system of charged particles, which could cause symmetry breaking. Different phases then appear, and a gap emerges in the region. By using fermionic probes and calculating the Green's function, the behavior of the modes around this gap can be studied.

\begin{figure}[ht!]
\centering
\includegraphics[scale=0.65]{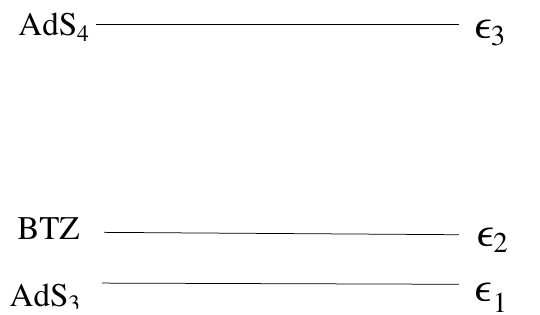}
\caption{Three limits around the black hole in the bulk, which can be seen by taking different orders of limits in $11d$.}
\label{fig:phases}
\end{figure}

By uplifting the theory to $11d$ and then considering the near-horizon limit with $q' = 0$, the near-horizon geometry decouples into $\mathrm{AdS}_4 \times \mathrm{M}^7$. In the near-horizon limit of a 3-charge black brane, the AdS part is effectively an $\mathrm{AdS}_3$. The decoupled regime in the bulk can be interpreted as a gap in the CFT, with its own degrees of freedom not communicating with those outside the gap.

If $q'$ is non-zero but small, then the black hole correspondingly has a small angular momentum, and the near-horizon geometry close to the horizon decouples into another sector, which would be an excited $\mathrm{AdS}_3$ or a rotating BTZ with energy $\epsilon_2$. In the UV, far from the horizon, the $11d$ geometry is $\mathrm{AdS}_4 \times \mathrm{S}^7$.

After uplifting the $4d$ theory to $5d$, depending on how the Killing spinors are fixed on the spherical part, the $5d$ spinors can be found. A notable fact is that the additional spatial dimension of the metric is not within the $\mathrm{AdS}$ section; it actually comes from the $\mathrm{S}^7$ part of the metric, which joins the $\mathrm{AdS}_2$ and together they build the near-horizon $\mathrm{AdS}_3$.

In future works, one could compute the central charge of the dual $\mathrm{CFT}_2$ and match the entropy of the deformed $\mathrm{AdS}_3$ sector to the Bekenstein–Hawking entropy of the 4D black hole using the Cardy formula. In addition, one could extend this analysis to include quantum corrections and compare with localization results in the ABJM theory. This study could also be carried out for various other supergravity models in different dimensions, and the nature of the gap and fermionic responses could be examined there as well.

\section*{Acknowledgements}
I would like to thank Mohammad Mahdi Sheikh-Jabbari, Mohammad Reza Mohammadi Mozaffar, Kazem Bitaghsir, Hesam Soltanpanahi, and Hajar Ebrahim for useful discussions.

 \medskip

\bibliography{fermi.bib}
\bibliographystyle{JHEP}
\end{document}